\documentclass[useAMS,usenatbib,usegraphicx]{mn2e}
\usepackage{times}

\newcommand{\HI}{{\sc H\,i}}
\newcommand{\itHI}{{H\,{\normalsize I}}}
\newcommand{\secHI}{H\,{\sevensize\bf I}}
\newcommand{\mJybeam}{mJy beam$^{-1}$}
\newcommand{\msun}{{$M_\odot$}}
\newcommand{\kms}{$\,$km$\,$s$^{-1}$}
\newcommand{\ltsima} {$\; \buildrel < \over \sim \;$}
\newcommand{\gtsima} {$\; \buildrel > \over \sim \;$}
\newcommand{\lta} {\lower.5ex\hbox{\ltsima}}
\newcommand{\gta} {\lower.5ex\hbox{\gtsima}}
\newcommand{\sauron}{{\textit {SAURON}}}
\newcommand{\miriad}{{\textit {MIRIAD}}}

\title[Early-type galaxies in different environments, an \itHI\ view]
{Early-type galaxies in different environments: an H\,{\Large\bf I} view}
\author[Tom Oosterloo et al.]  {Tom
  Oosterloo$^{1,2}$\thanks{E-mail:oosterloo@astron.nl}, Raffaella
  Morganti$^{1,2}$, Alison Crocker$^{3,4}$\thanks{ASTRON/JIVE Summer Student
    2007}, Eva J\"{u}tte$^{1,5}$,\and Michele Cappellari$^3$, Tim de
  Zeeuw$^{6,7}$, Davor Krajnovi\'c$^{3,6}$, Richard McDermid$^8$, \and Harald
  Kuntschner$^9$, Marc Sarzi$^{10}$, Anne-Marie Weijmans$^{7,11}$
  \\
  $^1$ ASTRON - Netherlands Institute for Radio Astronomy,
  Postbus 2, 7990 AA Dwingeloo, The Netherlands \\
  $^2$ Kapteyn Astronomical Institute, University of Groningen
  Postbus 800, 9700 AV Groningen, The Netherlands \\
  $^3$ Sub-department of Astrophysics, University of Oxford, Denys Wilkinson 
  Building, Keble Road, Oxford OX1 3RH \\
  $^4$ Department of Astronomy, University of Massachusetts, 710 North
  Pleasant Street
  Amherst, MA 01003-9305 USA\\
  $^5$ Astronomisches Institut, Ruhr-Universit\"{a}t Bochum,
  Universit\"{a}tsstrasse 150,
  D-44801 Bochum, Germany\\
  $^6$ European Southern Observatory, Karl-Schwarzschild Strasse 2,
  85748 Garching bei M\"{u}nchen, Germany \\
  $^7$ Sterrewacht Leiden, Universiteit Leiden, Niels Bohrweg 2, 2333 CA
  Leiden,
  The Netherlands\\
  $^8$ Gemini Observatory, 670 N.\ A'ohoku Place, Hilo, Hawaii 96720 USA      \\
  $^9$ Space Telescope European Coordinating Facility, Karl-Schwarzschild-Str.
  2,
  D-85748 Garching bei M\"{u}nchen, Germany\\
  $^{10}$ Centre for Astrophysics Research, Science \& Technology Research
  Institute, University of Hertfordshire, Hatfield, United Kingdom\\
  $^{11}$ Dunlap Institute for Astronomy \& Astrophysics, University of
  Toronto, 50 St. George Street, Toronto, ON M5S 3H4, Canada }

\begin{document}

\date{Accepted 2010 July 12. Received 2010 July 7 ; in original form 2010 March 5 }

\pagerange{\pageref{firstpage}--\pageref{lastpage}} \pubyear{2010}

\maketitle

\label{firstpage}

\begin{abstract} We present an analysis of deep Westerbork Synthesis Radio
  Telescope observations of the neutral hydrogen in 33 nearby early-type
  galaxies selected from a representative sample studied earlier at optical
  wavelengths with the \sauron\ integral-field spectrograph. This is the
  deepest homogeneous set of \HI\ imaging data available for this class of
  objects.  The sample covers both field environments and the Virgo cluster.
  Our analysis shows that gas accretion plays a role in the evolution of field
  early-type galaxies, but less so for those in clusters.

  The \HI\ properties of \sauron\ early-type galaxies strongly depend on
  environment.  For detection limits of a few times $10^6$ $M_\odot$, \HI\ is
  detected in about 2/3 of the field galaxies, while $<$10\% of the Virgo
  objects are detected. In about half of the detections, the \HI\ forms a
  regularly rotating disc or ring.  In many galaxies unsettled tails and
  clouds are seen.  All \HI\ discs have counterparts of ionised gas and inner
  \HI\ discs are also detected in molecular gas.  The cold ISM in the central
  regions is dominated by molecular gas ($M_{\rm H_2}/M_{\rm HI} \simeq 10$).
  Assuming our sample is representative, we conclude that accretion of \HI\ is
  very common for field early-type galaxies, but the amount of material
  involved is usually small and the effects on the host galaxy are, at most,
  subtle. Cluster galaxies appear not to accrete \HI, or the accreted material
  gets removed quickly by environmental effects.  The relation between \HI\
  and stellar population is complex. The few galaxies with a significant young
  sub-population all have inner gas discs, but for the remaining galaxies
  there is no trend between stellar population and \HI\ properties. A number
  of early-type galaxies are very gas rich, but only have an old population.
  The stellar populations of field galaxies are typically younger than those
  in Virgo.  This is likely related to differences in accretion history.
  There is no obvious overall relation between gas \HI\ content and global
  dynamical characteristics except that the fastest rotators all have an \HI\
  disk. This confirms that if fast and slow rotators are the result of
  different evolution paths, this is not strongly reflected in the {\sl
    current} \HI\ content.  In about 50\% of the galaxies we detect a central
  radio continuum source. In many objects this emission is from a
  low-luminosity AGN, in some it is consistent with the observed star
  formation. Galaxies with \HI\ in the central regions are more likely
  detected in continuum. This is due to a higher probability for star
  formation to occur in such galaxies and not to \HI-related AGN fuelling.
\end{abstract}

\begin{keywords} galaxies: elliptical and lenticular, cD --- galaxies:
evolution \end{keywords}

\section{Introduction} \label{sec:introduction}

One of the central topics in current extra-galactic astronomy is how
early-type galaxies form and how their properties change through cosmic times.
This is a particularly difficult task as this class of objects shows a
diversity in its properties that goes beyond present-day simulations.  The
generally accepted framework is that early-type galaxies form in a
hierarchical way through the accretion and merging of smaller systems.
Although hierarchical growth in itself is a relatively simple premise, the
details of early-type galaxy formation and evolution are very complex and
depends on a large number of parameters which is likely the reason for the
observed variety of the final galaxies.

One of the main issues is how much gas is involved in the accretion/merging
process and to what extent it involves approximately equal-mass systems or the
accretion of small companions. Theoretical work has shown that the amount of
gas involved in the growth of early-type galaxies can be a major factor, in
particular in determining the morphological and dynamical structure of
early-type galaxies.  For example, more anisotropic and slowly rotating
galaxies would result from predominantly collisionless major mergers, while
faster rotating galaxies are produced by more gas-rich mergers and accretions.
The inner structure (i.e.\ cores vs cusps) is also likely related to the type
of merger/accretion
\citep[e.g.][]{Bender92,Jesseit05,Naab06,Hopkins09,Jesseit09}.

Some arguments suggest that the evolution of early-type galaxies is 'dry',
i.e.\ gas does not play an important role. The basic argument is that the
amount of stars in red galaxies has at least doubled since $z=1$, while the
red colours of early-type galaxies indicate that they are dominated by old
stars. This indicates that the growth since $z=1$ is dry, i.e.\ it is not
accompanied with much star formation \citep[e.g.][]{Bell04,Dokkum05,Tal09}.

While this may suggest that globally the amount of gas involved since $z=1$ is
at most modest, several observational studies, touching on several topics,
show that gas does play at least some role. For example, stellar-population
studies show that many systems do contain a (often small) sub-population of
relatively young stars that may have formed from accreted gas
\citep[e.g.][]{Trager00,Tadhunter05,Yi05, Serra06,Serra08,Kaviraj10}.
Similarly, early work by, e.g., \citet{Malin83} and \citet{Schweizer92}, has
shown that direct morphological signs of accretion are observed in a large
fraction of early-type galaxies and that such signs correlate with the
presence of a young stellar sub-population, indicating the presence of gas in
these accretions.  More recent work has shown that this is also the case for
samples of early-type galaxies that originally seemed to support the
dry-merging hypothesis \citep{Donovan07,Sanchez09,Serra10}. Dynamically
distinct stellar and gaseous sub-components are often found in early-type
galaxies. In many cases, such sub-components are both chemically and
kinematically distinct \citep{McDermid06}, strongly suggesting that external
gas has entered the system. The orbital structure of fast-rotating early-type
galaxies also seems to indicate that gas was involved in their evolution
\citep{Emsellem07,Cappellari07}.  Finally, the tight scaling relations found
for early-type galaxies place a rather conservative upper limit on the
fraction of stellar mass assembled via dissipationless merging
\citep[e.g.][]{Nipoti03,Nipoti09}.

Recently, also {\em direct evidence} for the importance of gas has been found.
Early-type galaxies in the nearby Universe used to be generally perceived to
be gas poor. Although indeed they typically have less cold gas than spiral
galaxies, it is now becoming clear that cold gas is present perhaps most of
them, in particular those in the field.  For example, for a detection limit of
a few $\times 10^7$ $M_\odot$, molecular gas is detected in at least a quarter
of early-type galaxies \citep{Welch03,Sage07,Combes07,Young10}.  Recent work
on the neutral hydrogen in early-type galaxies suggests that, in terms of
detection limits of a few times $10^6$ $M_\odot$, about half the field
early-type galaxies are detected \citep{Morganti06,Grossi09,Serra09}. The \HI\
datacubes obtained by \citet{Morganti06} also show, in terms of the
characteristics of the neutral hydrogen detected, the class of field
early-type galaxies appears to be rich and varied, much more so than spirals.

An interesting aspect is that early-type galaxies in clusters appear to have
different gas properties from those in the field \citep{diSerego07,Serra09}.
Therefore, if gas plays a role in the evolution of early-type galaxies, this
should become visible by comparing properties of cluster early-type galaxies
with those of objects in the field.

In this paper we expand on the results obtained in \citet{Morganti06} with
particular focus on the effect of environment. We present deep \HI\ imaging
observations for 22 galaxies selected from the \sauron\ sample
\citep{deZeeuw02} where we have, in contrast to \citet{Morganti06}, also
selected galaxies from the Virgo cluster.  Combining these new data with those
of \citet{Morganti06} resulted in deep \HI\ data for 33 \sauron\ galaxies
north of declination +10$^\circ$.  This is the largest and deepest collection
of \HI\ imaging data available for early-type galaxies. The \sauron\ sample is
well suited for a detailed comparison between field and cluster galaxies: for
all \sauron\ galaxies a wealth of information is available, including 3-D
spectroscopy of the stars and of the ionised gas as well as data obtained in
many other wavebands. A potential concern is that in principle the \HI\
properties were part of the selection of the \sauron\ sample. However, in
practise this affected the selection of only a few objects of the \sauron\
sample of 48 early-type galaxies. Importantly, the selection was done
independent of environment.  Therefore, results based on a comparison of the
\HI\ properties of \sauron\ galaxies in different environments should be
robust.

The paper is organised in the following way. We describe the sample selection
and the observations in Section\,\ref{sec:observations}. In
Section\,\ref{sec:results} we present the results for the new \HI\ detected
galaxies. In Section\,\ref{sec:discussion} we discuss the observed \HI\
properties of early-type galaxies in the context of their evolution.

\begin{table*}
\tabcolsep=3pt
\begin{tabular}{ccccccccccc}
\hline\hline
NGC  &  $V_{\rm centr}$ & D &pc/$^{\prime\prime}$ &Date & Int.Time & Beam & Noise \HI\ & Noise Cont. & \HI\ contours \\
     &            \kms\   & Mpc  &                      &     &   h      & $^{\prime\prime}\times ^{\prime\prime} (^\circ)$
                                           & \mJybeam\  & \mJybeam\   & $10^{19}$ cm$^{-2}$       \\
(1)  &    (2)  & (3)  & (4)          & (5) & (6)      & (7)  & (8)
& (9)    & (10)      \\
\hline
524  &  2379 & 23.3 & 113 &   & $3\!\times\!12$ & $85\!\times\!23$   & 0.28 & 0.062 & 1,2  \\
821  &  1742 & 23.4 & 113 &   & $4\!\times\!12$ & $72\!\times\!13$   & 0.20 & 0.050 &  --\\
3032 &  1561 & 21.4 & 104 &   & $4\!\times\!12$ & $43\!\times\!25$   & 0.19 & 0.048 & 2, 5, 10, 20, 50 \\
3377 &   698 & 10.9 &  53 &   & 12              & $73\!\times\!24$   & 0.37 & 0.053 &  --\\
3379 &   877 & 10.3 &  50 &   & 12              & $73\!\times\!24$   &      & 0.052 &  --\\
3384 &   729 & 11.3 &  55 &   & 12              & $73\!\times\!24$   &      & 0.052 &  1, 2, 5\\
3489 &   688 & 11.7 &  57 &   & $4\!\times\!12$ & $76\!\times\!25$   & 0.20 & 0.047 &  0.5, 1, 2, 5, 10 \\
3608 &  1201 & 22.3 & 108 &   & $4\!\times\!12$ & $62\!\times\!22$   & 0.22 & 0.043 &  2, 5 \\
4262 &  1361 & 15.4 &  75 &   & 12              & $68\!\times\!24$   & 0.42 & 0.069 &  10, 20, 50\\
4374 &  1016 & 18.5 &  90 &   & 12              & $59\!\times\!26$   & 0.72 &       &  --\\
4382 &   745 & 17.9 &  87 &   & 12              & $59\!\times\!26$   & 0.39 & 0.062 & -- \\
4387 &   550 & 17.9 &  87 &   & 12              & $67\!\times\!24$   & 0.45 & 0.490 &  --\\
4458 &   676 & 16.4 &  79 &   & 12              & $71\!\times\!25$   & 0.41 & 0.128 &  --\\
4459 &  1182 & 16.1 &  78 &   & 12              & $69\!\times\!25$   & 0.40 & 0.065 &  --\\
4473 &  2210 & 15.3 & s 74 &   & 12              & $71\!\times\!25$   & 0.42 & 0.151 &  --\\
4477 &  1327 & 16.5 &  80 &   & 12              & $71\!\times\!25$   & 0.38 & 0.124 &  --\\
4486 &  1272 & 17.2 &  83 &   & 12              &  --                & --   & --   & --\\
4550 &   407 & 15.5 &  75 &   & 12              & $74\!\times\!25$   & 0.39 & 0.121 &  --\\
4552 &   288 & 15.8 &  77 &   & 12              & $78\!\times\!24$   & 0.60 & 0.098 &  --\\
4564 &  1116 & 15.8 &  77 &   & 12              & $76\!\times\!25$   & 0.68 & 0.144 &  --\\
4621 &   431 & 14.9 &  72 &   & 12              & $76\!\times\!25$   & 0.39 & 0.073 &  --\\
4660 &  1082 & 15.0 &  73 &   & 12              & $77\!\times\!25$   & 0.39 & 0.059 &  --\\
\hline
\end{tabular}
\caption{Summary of the observations of the galaxies in the sample.
  (1) Galaxy identifier. (2) Systemic velocity at which we centred the
  \HI\ observation band.  (3) Galaxy distance \citep[][corrected by subtracting 0.06 mag, 
  see \citet{Mei05}]{Tonry01}, \citet{Tully88} or from the LEDA database
assuming a Hubble flow with $H_\circ = 75$ km
  s$^{-1}$ Mpc$^{-1}$. (4) Linear scale. (5) Date of observation. (6)
  Integration time in hours. (7) Beam size. (8) Noise level in the \HI\ cube (natural). (9)
  Noise level of the continuum image. (10) Contour levels of the total
  intensity images shown in Fig.~\ref{fig:himaps}. 
\label{tab:table1}}
\end{table*}

\section{Sample selection and \secHI\  observations} 
\label{sec:observations} 

We observed 22 galaxies from the \sauron\ sample \citep{deZeeuw02} with the
Westerbork Synthesis Radio Telescope (WSRT).  For one of these galaxies
(NGC~4486/M87), the very strong radio continuum emission prevented us to
obtain a data cube of good enough quality. This object is, therefore, excluded
from the analysis.  As mentioned above, the newly observed galaxies expand the
sample presented in \citet{Morganti06} with objects down to the declination
limit of $+10^\circ$ and including galaxies that are member of the Virgo
cluster. Lowering the declination limit is a compromise between increasing the
sample size while maintaining good image quality. For galaxies close to the
declination limit, due to the east-west layout of the WSRT, the beam
elongation becomes significant, and spatially resolved information becomes
limited.  However, given that most of the \HI\ structures detected so far are
very extended and of low surface brightness, the large WSRT beam is not a
major disadvantage.  For the observations, a single observing band of 20~MHz
(corresponding to $\sim$ 4000\,kms$^{-1}$), centred on the systemic velocity
of the target, and 1024 channels for both polarisations was used. All
observations were done with the maxi-short antenna configuration. In most
cases, we observed the targets for 12 h. For five objects that turned out to
have interesting but faint \HI\ detections, or for which the single 12-h
observation gave a tentative detection, follow up observations ($3 \times
12$h) were obtained. The details of the observations are given in
Table\,\ref{tab:table1}.

The calibration and analysis were done using the \miriad\ package
\citep{Sault95}. The data cubes were constructed with a robust weighting equal
to 0 \citep{Briggs95}.  In the cases where faint, extended structures were
detected (e.g.\ NGC~3489), additional cubes were constructed using natural
weighting. The cubes were made by averaging channels in groups of two,
followed by Hanning smoothing, resulting in a velocity resolution of 16 \kms.
This was done to match the spectral resolution to the expected line widths.
The r.m.s.\ noise and restoring beam sizes of each cube are given in
Table\,\ref{tab:table1}. For reference: a beam size of 1 arcminute corresponds
to about 5 kpc for a galaxy at a distance of 15 Mpc.

The line-free channels were used to obtain an image of the radio continuum of
each galaxy. The continuum images were made with uniform weighting. The
r.m.s.\ noise and beam of these images are also given in
Table\,\ref{tab:table1}.  Radio continuum emission was not detected in 14 of
the objects. All detected continuum sources are unresolved.

\begin{table*}
\tabcolsep=12pt
\begin{tabular}{cccccccccc}
\hline\hline
NGC     & Type &   $M_{\rm HI}$       &$M_{\rm HI}/L_B$ & \HI\ morph &
Env &S$_{\rm 1.4GHz}$ & $\log P_{\rm 1.4GHz}$\\
             &          &   $M_\odot$         &           &          & & mJy     &  W/Hz   &      \\
(1)     &(2) &  (3)       & (4)               & (5)       & (6)    &  (7)& (8)      \\
\hline
 524   & S0 &   $2.6\times 10^6$   & 0.000045     & C & F & 2.19    &  20.13   \\
 821   & E  &   \llap{$<$}$4.4\times 10^6$  & \llap{$<$}$0.00018$   & N & F & \llap{$<$}$0.15$ & \llap{$<$}18.97 \\
3032   & S0 &   $9.6\times 10^7$   & 0.019        & D & F & 5.66    & 20.47    \\
3377   & E  &   \llap{$<$}$1.8\times 10^6$  & \llap{$<$}$0.00023$   & N & F & \llap{$<$}$0.16$  & \llap{$<$}18.33 \\
3379   & E  &   \llap{$<$}$1.8\times 10^6$  & \llap{$<$}$0.0001$    & N & F & 0.87    & 19.02    \\
3384   & S0 &   $1.2\times 10^7$   & 0.00116      & C & F & \llap{$<$}$0.16$ & \llap{$<$}18.35 \\
3489   & S0 &   $5.8\times 10^6$   & 0.00007      & D & F & 1.22    & 19.28    \\
3608   & E  &   $2.5\times 10^6$   & 0.00024      & C & F & 0.5     & 19.45    \\ 
4262   & S0 &   $6.4\times 10^8$   & 0.115        & D & C & 0.71    & 19.28    \\
4374   & E  &   \llap{$<$}$9.3\times 10^6$  & \llap{$<$}$0.00019$   & N & C & 1500&  22.8   &          \\
4382   & S0 &   \llap{$<$}$5.0\times 10^6$  & \llap{$<$}$0.0001$    & N & C & \llap{$<$}$0.19$ & \llap{$<$}18.83 \\
4387   & E  &   \llap{$<$}$1.2\times 10^7$  & \llap{$<$}$0.0037$    & N & C & \llap{$<$}$1.47$ & \llap{$<$}19.73 \\
4458   & E  &   \llap{$<$}$4.6\times 10^6$  & \llap{$<$}$0.00126$   & N & C & \llap{$<$}$0.38$ & \llap{$<$}19.07 \\
4459   & S0 &   \llap{$<$}$3.9\times 10^6$  & \llap{$<$}$0.00025$   & N & C & 1.52    & 19.65    \\
4473   & E  &   \llap{$<$}$3.9\times 10^6$  & \llap{$<$}$0.0002$    & N & C & \llap{$<$}$0.45$ & \llap{$<$}19.08 \\
4477   & S0 &   \llap{$<$}$4.1\times 10^6$  & \llap{$<$}$0.00027$   & N & C & 1.16    & 19.55    \\
4550   & S0 &   \llap{$<$}$3.7\times 10^6$  & \llap{$<$}$0.0007$    & N & C & \llap{$<$}$0.36$ & \llap{$<$}18.85 \\
4552   & E  &   \llap{$<$}$5.3\times 10^6$  & \llap{$<$}$0.00020$   & N & C & 84.1    & 21.38    \\
4564   & E  &   \llap{$<$}$5.8\times 10^6$  & \llap{$<$}$0.00065$   & N & C & \llap{$<$}$0.43$ & \llap{$<$}19.09 \\
4621   & E  &   \llap{$<$}$4.9\times 10^6$  & \llap{$<$}$0.00018$   & N & C & \llap{$<$}$0.22$ & \llap{$<$}18.74 \\
4660   & E  &   \llap{$<$}$2.4\times 10^6$  & \llap{$<$}$0.00031$   & N & C & \llap{$<$}$0.18$ & \llap{$<$}18.65 \\
\hline
1023   & S0 &   $2.1\times 10^9$   &  0.046       & A & F & 0.4     & 18.7     \\
2549   & S0 &   \llap{$<$}$2.0\times 10^6$  &  \llap{$<$}$0.00043$  & N & F & \llap{$<$}$0.1$  & \llap{$<$}$18.3$  \\
2685   & S0 &   $1.8\times10^9$    &  0.27        & D & F & 2       & 19.8     \\
2768   & E  &   $1.7\times10^8$    &  0.0038      & A & F & 10.9    & 20.8     \\
3414   & S0 &   $1.6\times 10^8$   &  0.0096      & D & F & 5.0     & 20.6     \\
4150   & S0 &   $2.5\times 10^6$   &  0.00078     & D & F & 0.8     & 19.2     \\
4278   & E  &   $6.9\times 10^8$   &  0.039       & D & F & 336.5   & 22.0     \\
5198   & E  &   $6.8\times 10^8$   &  0.030       & A & F & 2.4     & 20.6     \\
5308   & S0 &   \llap{$<$}$1.5\times 10^7$  &  \llap{$<$}$00064$    & N & F & \llap{$<$}$0.24$ & \llap{$<$}$19.5$  \\
5982   & E  &   $3.4\times 10^7$   &  0.00060     & C & F & 0.5     & 20.1     \\
7332   & S0 &   $6.0\times 10^6$   &  0.00038     & C & F & \llap{$<$}$0.13$ & \llap{$<$}$18.9$  \\
7457   & S0 &   \llap{$<$}$2.0\times10^6$   &  \llap{$<$}$0.00032$  & N & F & \llap{$<$}$0.11$ & \llap{$<$}$18.5$  \\
\hline 
\end{tabular}
\caption{Measurements based on our radio
observations. The top part of this table is based on the observations
presented here. For completeness, we include the parameters from
\citet{Morganti06}. 
(1) Galaxy identifier. (2) Hubble type (NED). (3) Total \HI\
mass. (4) Ratio of total \HI\ mass and the absolute $B$-band luminosity
$L_B$.  (5) Code describing \HI\ morphology: N: not detected, C: isolated
cloud, A: accretion, D: disc (6) Environment code: F: field, C: Virgo cluster
(7) Continuum flux (or $3\sigma$ upper limits) at 1.4 GHz. (8) Total radio
power at 1.4 GHz.
\label{tab:table2}}
\end{table*}

\begin{figure*} 
\includegraphics[width=7cm]{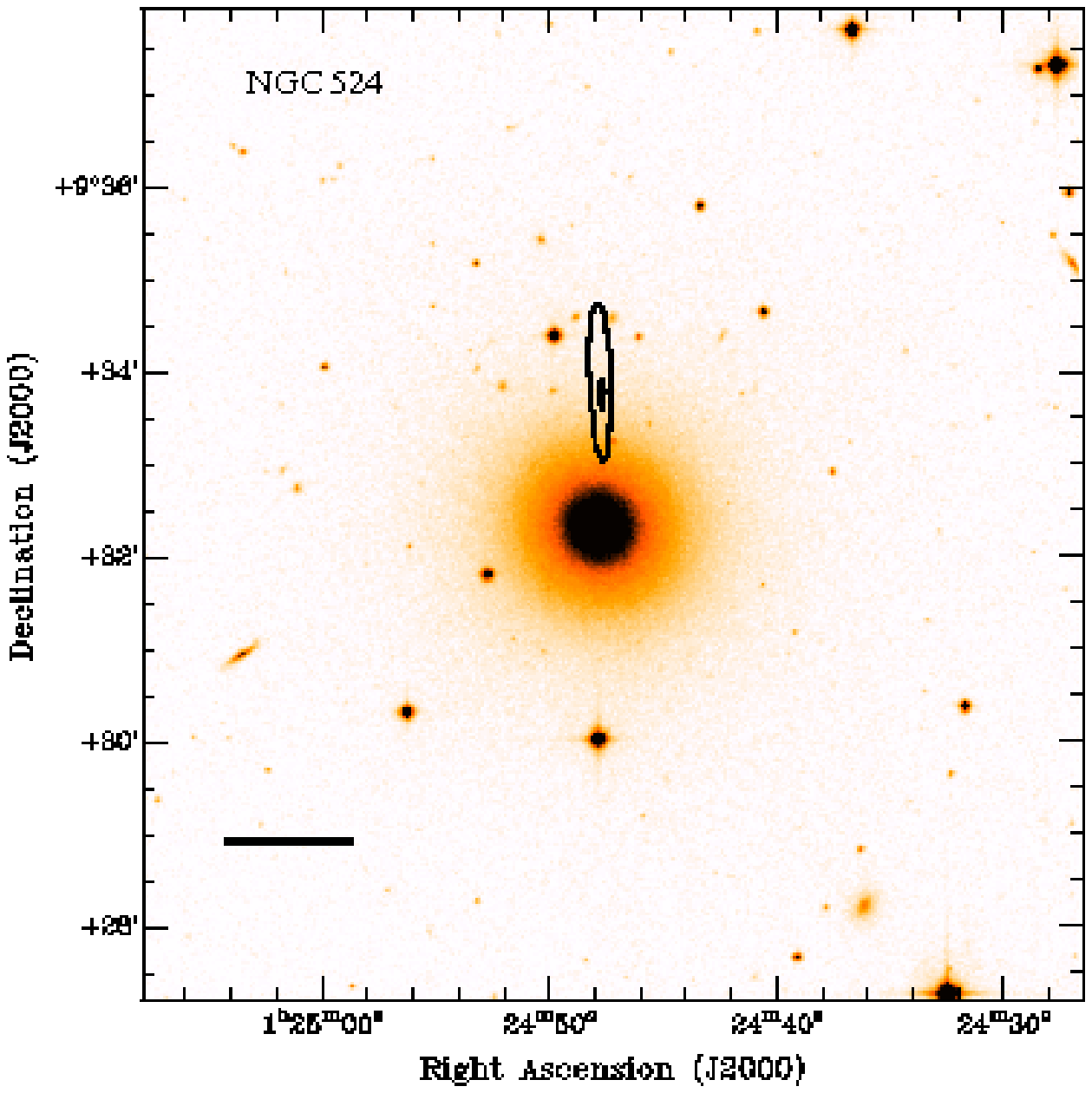}
\includegraphics[width=7cm]{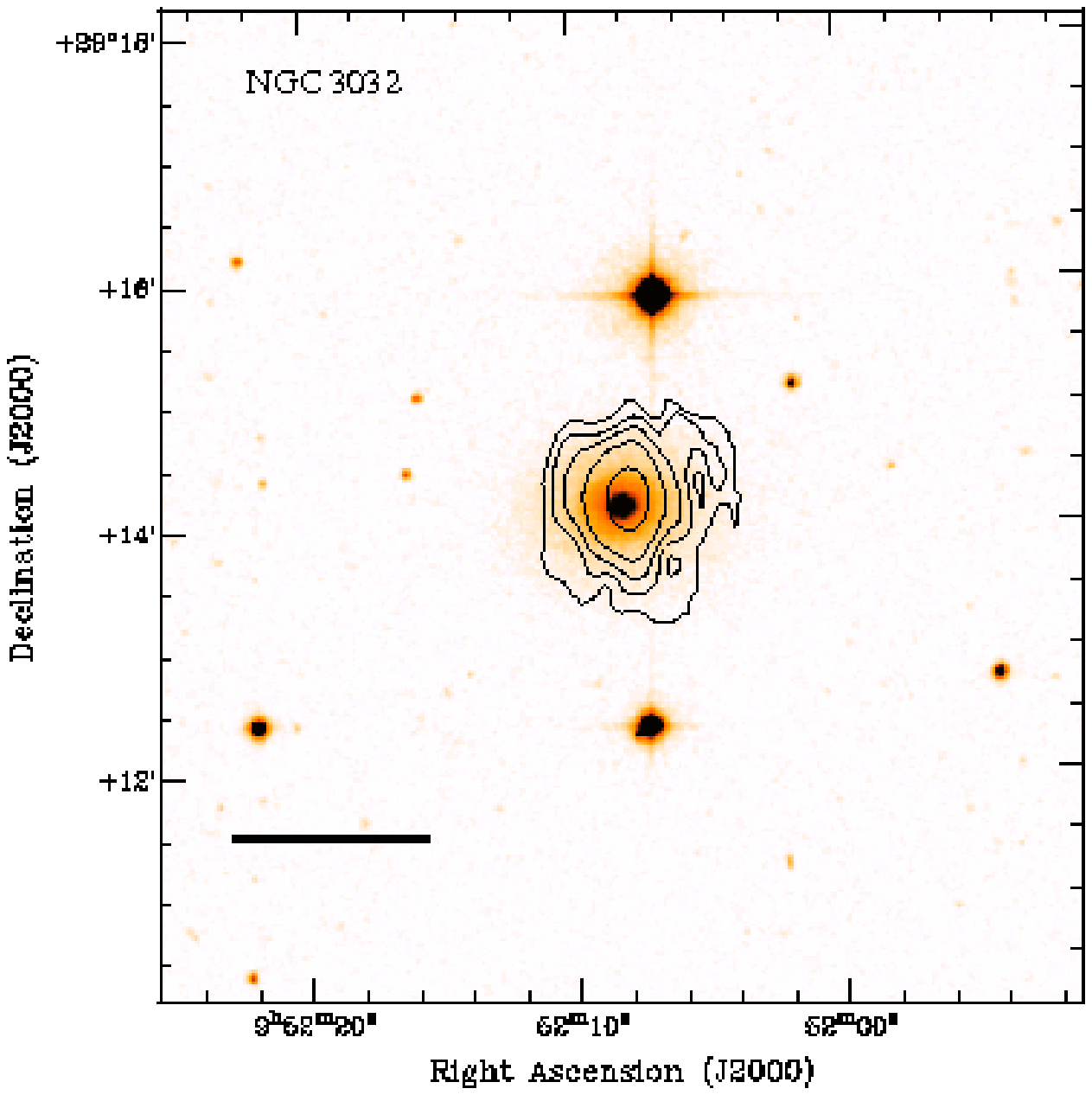}\\
\includegraphics[angle=270,width=7cm]{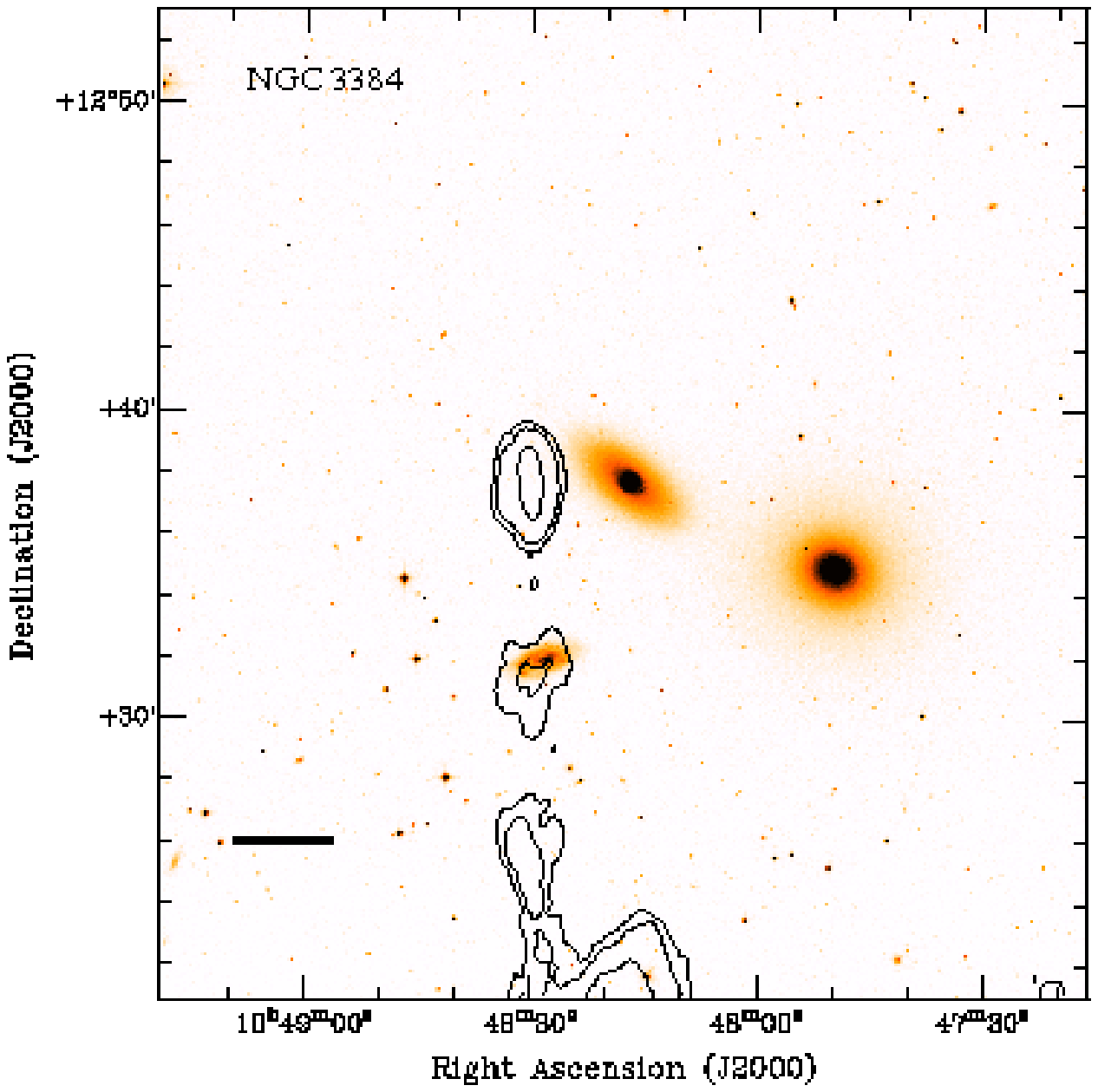}
\includegraphics[angle=270,width=7cm]{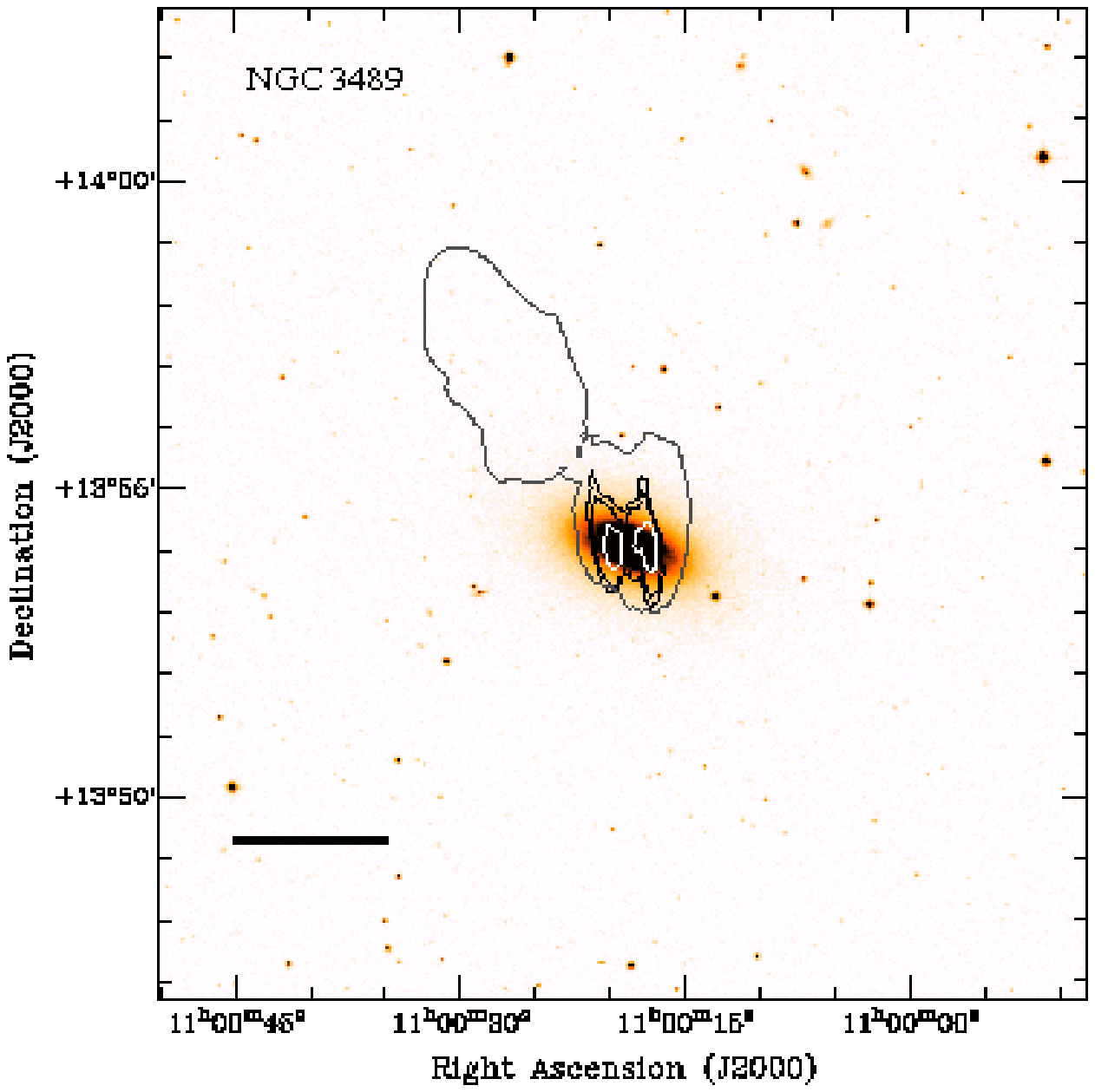} \\
\includegraphics[width=7cm]{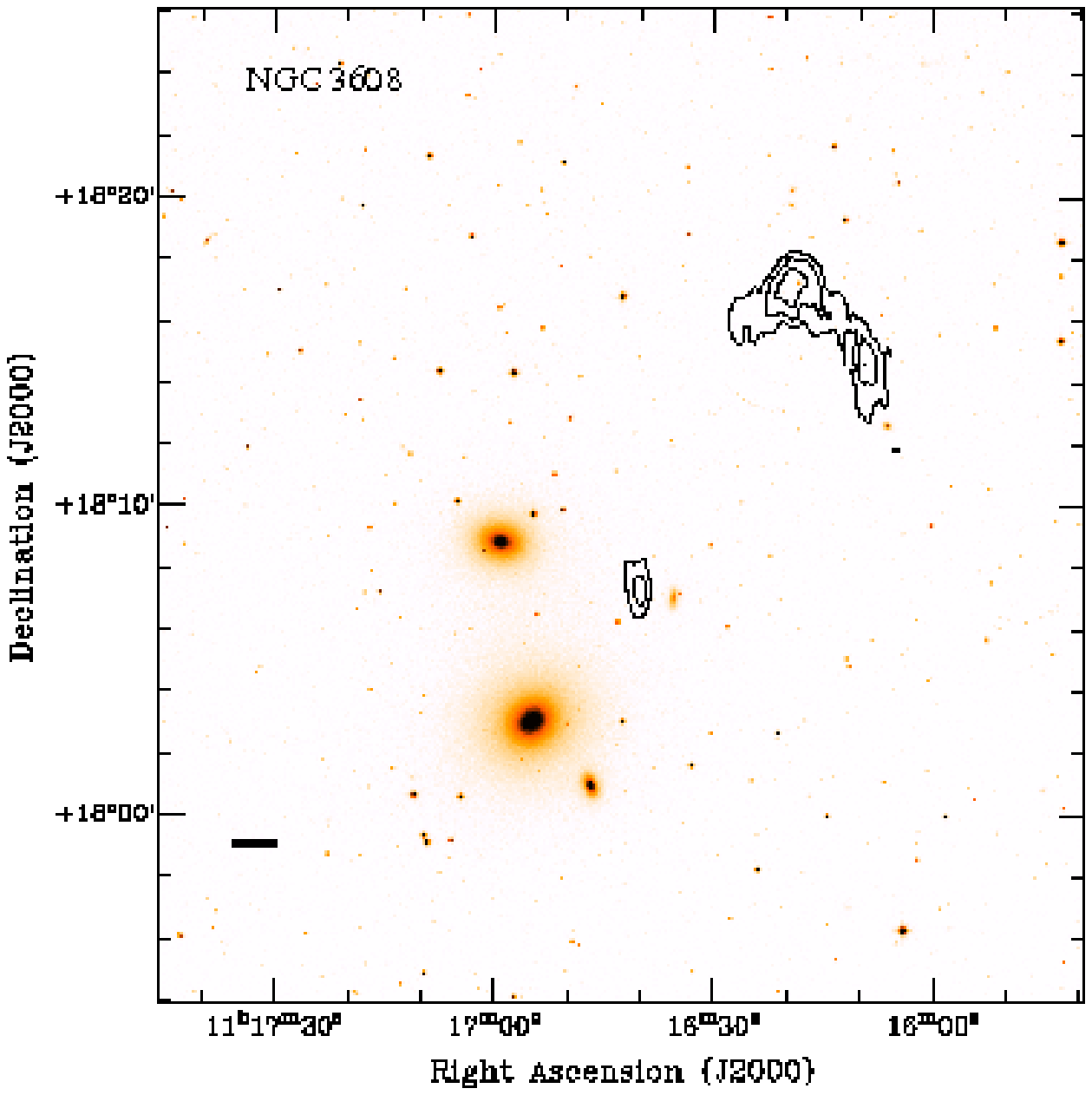}
\includegraphics[width=7cm]{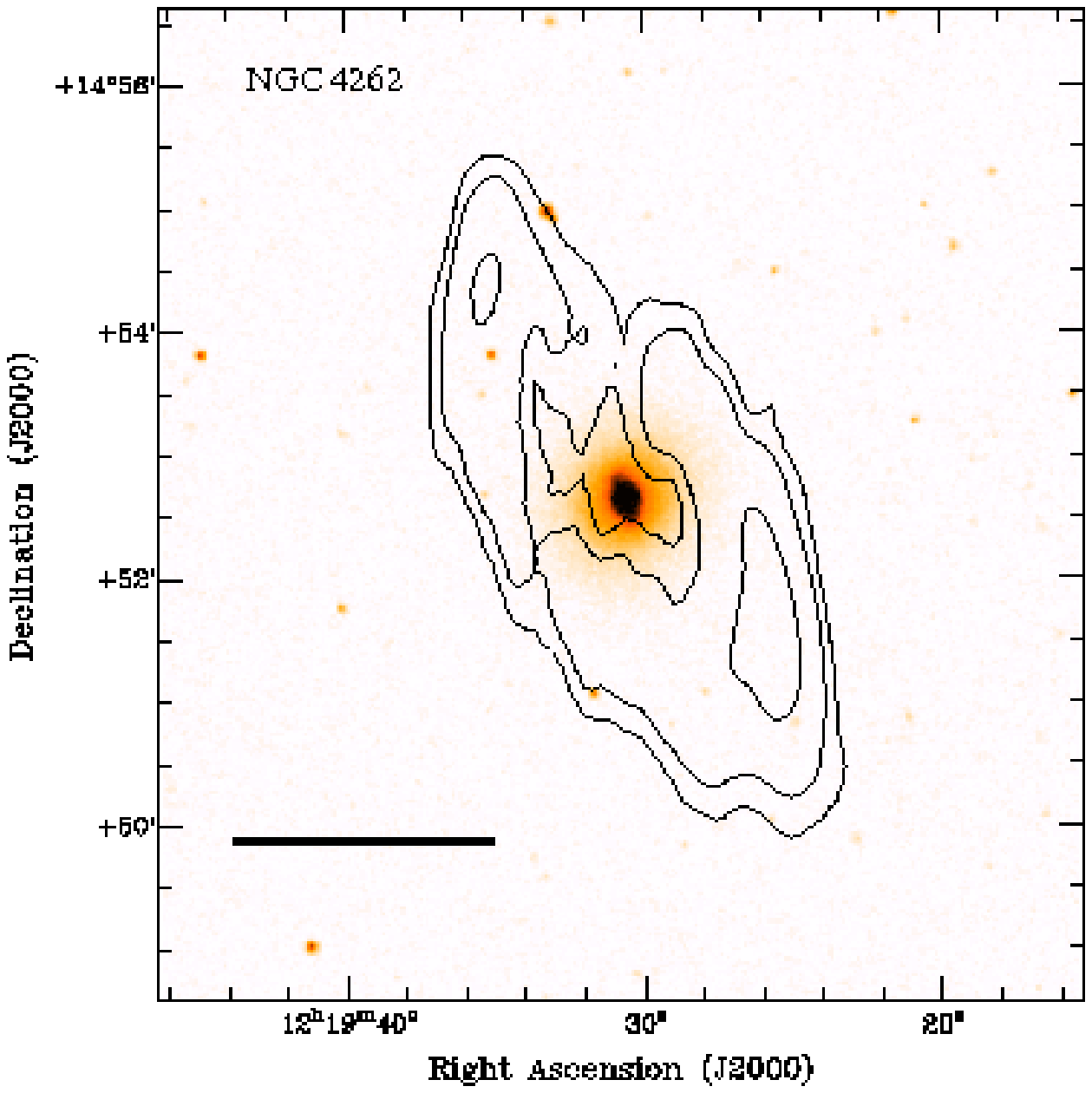} \caption{Total \HI\ intensity
  images (contours) superimposed onto Digital Sky Survey optical images of the
  newly detected objects - NGC~524, NGC~3032, NGC~3384 (part of the Leo Ring),
  NGC~3489, NGC~3608 and NGC~4262. The contour levels are given in
  Table~\ref{tab:table1}. The horizontal bar in each panel indicates 10 kpc.
  Total \HI\ intensity image for NGC~3384 has been made excluding the
  high-velocity system belonging to <NGC 3389 (see text) \label{fig:himaps}}
\end{figure*}

\section{Results} \label{sec:results}

Neutral hydrogen is detected in our new observations in, or near, six
galaxies.  Some details about the individual detections are given at the end
of this section. Five of the detections are of \HI\ in field early-type
galaxies (out of 8 field galaxies observed), one is in a galaxy that is a
member of the Virgo cluster (out of 13 Virgo galaxies observed).  Figure
\ref{fig:himaps} shows the \HI\ total intensity images of the detections.  In
the strongest detections - NGC~3032, NGC~3489, NGC~4262 - the \HI\ is located
in a regular disc/ring-like structure, which in the case of NGC~3489 connects
to a long low-surface brightness tail of \HI.

In NGC~3608, the \HI\ is detected about $\sim 12$ arcmin (corresponding to
$\sim 70$ kpc) from the galaxy, at a velocity close to the systemic velocity
of NGC 3608 and without any obvious optical counterpart. A few other (also
early-type) galaxies are nearby so it is difficult to assign the \HI\ to NGC
3608 unambiguously.  A fifth \HI\ detection is NGC~3384. Together with
NGC~3377 and NGC~3389, this galaxy is surrounded by the well-known Leo Ring
\citep{Schneider83,Schneider89} of which we conclude that at least one \HI\
cloud is likely associated with NGC 3384.  Finally, in NGC 524 we detect a
small \HI\ cloud near the edge of the optical body. The results for the field
galaxies, with regard to both morphology and detection rate, are in line with
those we obtained in our previous study of \sauron\ galaxies
\citep{Morganti06}.

A clear result is that Virgo early-type galaxies clearly have different \HI\
properties than field galaxies. This is discussed in more detail in Secs
\ref{sec:environment} and \ref{sec:stellarPop}.  For galaxies where no \HI\
was detected, the upper limits on the \HI\ mass were calculated as three times
the statistical error of a signal with a width of 200 \kms\ over one
synthesised beam. The upper limits of the \HI\ mass range from a few times
$10^6$ $ M_\odot$ to $3\times10^7$ $ M_\odot$. The quantitative results are
given in Table \ref{tab:table2}. Since we will discuss the \HI\ results for
all \sauron\ galaxies observed, we also include the parameters of the galaxies
observed by \citet{Morganti06}.

The \HI\ masses of the newly detected objects range between $10^7\ M_\odot$ to
a few times $10^8\ M_\odot$. The relative gas content ($M_{\rm HI}/L_{\rm B}$)
ranges from $< 0.0001\ M_\odot/L_\odot$ to $0.115\ M_\odot/L_\odot$. Typical
values for $M_{\rm HI}/L_{\rm B}$ for spiral galaxies, depending on type and
luminosity, range from 0.1 $M_\odot/L_\odot$ to above 1.0 $M_\odot/L_\odot$
\citep{Roberts94}. These values underline the well-known fact that early-type
galaxies are \HI\ poor relative to spiral galaxies. The sizes of the \HI\
structures observed in this sample vary between a few kpc up to $\sim$40 kpc.
Figure \ref{fig:himaps} shows that the peak column density istypically at most
a few times 10$^{20}$ cm$^{-2}$. As already found in many earlier studies
\citep[e.g.][]{Driel91,Morganti97,Morganti06,Serra06,Oosterloo07}, these
column densities are lower than the critical surface density for star
formation \citep{Kennicutt89,Schaye04,Bigiel08}. Although this result implies
the absence of widespread star formation activity, given the relatively low
spatial resolution of our data, the column density will be above the star
formation threshold in local, small regions and some star formation can be
expected.

Before we further discuss the possible implications of our \HI\ observations, 
we summarise the \HI\ characteristics for the individual objects detected.

{\sl NGC~524} - We detect a small \HI\ cloud in the outer regions of this
galaxy. Possibly this corresponds to a small gas-rich companion, although no
direct optical counterpart is visible on fairly deep optical images
\citep{Jeong09}. A small galaxy, of which the redshift is not known, is seen
about 1 arcmin from the \HI\ cloud. Possibly a small companion is stripped
from its \HI\ by NGC 524, in a similar way as is occurring in, e.g., NGC 4472
\citep{McNamara94}. NGC 524 has a weak disc of ionised gas which rotates with
the same sense as the stars \citep{Sarzi06}. This disc is also detected in CO
with an implied molecular gas mass of $1.6\times10^8$ $M_\odot$ (Crocker et
al.\ unpublished). No counterpart to this gas disc is detected in \HI,
implying that most of the cold ISM in the central regions of NGC 524 is in the
form of molecular gas.

{\sl NGC\,3032} - In this galaxy, a small regularly rotating \HI\ disc is
found, with a total \HI\ mass of $9.7 \times 10^7$\,\msun. The \HI\ disc
co-rotates with the ionised gas observed \citep{Sarzi06}, although the detail
in the observed velocity field of the ionised gas is limited by binning
effects. The molecular gas is found in a centrally concentrated rotating
structure \citep{Young08}. The ionised, molecular and neutral gas are
co-rotating.  Interestingly, all these gaseous components are {\em
  counter-rotating} with respect to the bulk of the stars in this galaxy,
strongly suggesting an external origin of the gas observed. Some stars have
formed from this gas disc because \citet{McDermid07} have found the presence
of a small stellar core that is counter-rotating to the bulk of the stellar
body (and hence co-rotating with the gas disc). The CO observations reveal a
molecular gas reservoir of $2.5$-$5.0 \times 10^8$\msun
\citep{Combes07,Young08}. Thus, also in NGC 3032 most of the cold ISM is in
the molecular phase.

{\sl NGC\,3377/3379/3384--The Leo Group} - The situation in NGC 3384 is very
complex. This galaxy is member of the M96 group that is famous for its large
\HI\ ring encircling several galaxies of this group \citep[the Leo
Ring;][]{Schneider83}. Subsequent to the observations reported here, we have
imaged the \HI\ over the entire region and a full study of the Leo Ring will
be published elsewhere. One result of this work is that the Leo Ring appears
to form a large spiral-like structure that appears to end very close, both in
space and in velocity, to NGC 3384. At the endpoint of this spiral-like
structure a fairly bright \HI\ cloud is observed and this is the cloud we
identify here with NGC 3384. However, it is conceivable that most of the
\HI\ of the Leo Ring originated from NGC 3384 \citep[see][]{Dansac10}.  A complication is that the
\HI\ spiral appears to be interrupted by an interaction of the spiral galaxy
NGC 3389 with the Leo Ring, modifying its structure.  NGC 3389 has a long tail
of \HI\ observed at velocities about 600 \kms\ redshifted with respect to the
\HI\ shown in Fig. \ref{fig:himaps}, but shows no optical signs of a tidal
interaction.

{\sl NGC\,3489} - In NGC\,3489 we find an inner rotating \HI\ structure aligned
with the galaxy, as well as a low-column density tail. The total \HI\ mass of
this barred galaxy is $5.8 \times 10^6$\,\msun. This galaxy  might be in the
process of accreting a small gas cloud or companion, and forming an inner disc
from this material. The central \HI\ structure shows  regular rotation with the
same sense of rotation as the ionised gas and the stellar component. Molecular
gas with a mass of $1.2 \times 10^7$\,\msun has been detected in this galaxy by
\cite{Combes07}.

{\sl NGC\,3608} - In the vicinity of NGC 3608 we detect a large \HI\
structure, about 40 kpc in size and about 70 kpc from the galaxy. No \HI\ was
detected on NGC 3608 itself. A smaller \HI\ cloud is also detected, closer to
NGC 3608.  Since the galaxy is a member of a loose group, the intergalactic
\HI\ clouds might reflect past interactions between the group members, but
none of the nearby group galaxies hosts \HI\ today. This field shows
similarities with the elliptical galaxy NGC~1490 \citep{Oosterloo04} where a
number of large \HI\ clouds (with a total \HI\ mass of almost $10^{10}$ \msun)
are observed that are lying along an arc 500 kpc in length and at a distance
of 100 kpc from NGC 1490. The stars and the ionised gas in NGC 3608 are
kinematically decoupled, at least in the centre, where the stars show a
regular rotation pattern.  No CO was found in this galaxy in the survey of
\citet{Combes07}.

{\sl NGC\,4262} - This strongly barred galaxy is the only object in a dense
environment in which we detect in \HI.  The cold gas is distributed in a large
ring.   This ring shows regular rotation, albeit with signs of non-circular
orbits. The observed structure is most likely a resonance ring due to the bar.
The ionised gas rotates in the same sense as  the \HI\, whereas the stellar
rotation is decoupled from that. No CO was detected by \cite{Combes07}.

\begin{figure} 
\centerline{\includegraphics[width=7cm]{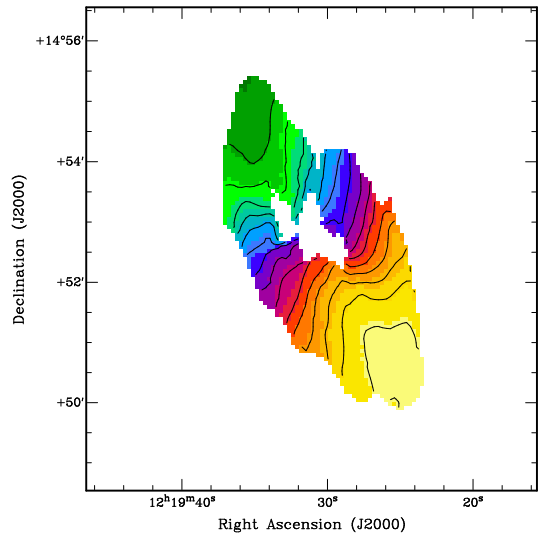}}
\caption{\HI\
velocity field of NGC~4262. Iso-velocity contours run from 1160 km s$^{-1}$
(top  left) to 1510 km s$^{-1}$ (bottom right) in steps of 25  km s$^{-1}$
\label{fig:xvDiagrams}} 
\end{figure}

\section{Discussion} \label{sec:discussion} Adding together the data from
Morganti et al.\ (2006) and the present work, we have data cubes of 33
early-type galaxies from the representative \sauron\ sample of 48. Using this
sample, we expand the analyses done in Morganti et al.  (2006). In particular,
we investigate here also the relation with the CO observations that have
become recently available as well as the effect of environment on the presence
of neutral hydrogen.  We discuss the effect of the environment in Sec. 4.1
while in the rest of the discussion we focus on field galaxies, where the
majority of the \HI\ detections are found.

\subsection{Environment} \label{sec:environment}

It is well established that the \HI\ properties of spirals in clusters are
strongly affected by the dense environment. Overall, spiral galaxies in
clusters are deficient in \HI\ compared to field spirals \citep[e.g.][and refs
therein]{Giovanardi83,Solanes01}, while imaging studies show that the \HI\
discs in spirals in clusters are clearly affected by the dense environment
\citep{Cayatte90,Oosterloo05,Chung09}. The above studies have shown that
dynamical interactions between galaxies, as well as stripping by the hot
intracluster medium, are responsible for removing a large fraction of the
neutral gas from cluster spirals. These mechanisms affect mainly the outer
disc regions, because the inner gas discs in cluster galaxies have similar
properties as those in field galaxies \citep{Kenney88,Young10}.  Another
important fact is that in the Virgo cluster the population of small, gas-rich
galaxies that can be accreted by larger galaxies in smaller than in the 
field \citep{Kent10}.  
Therefore, galaxies entering a cluster not only lose gas, but
they are also not able to replenish their gas supply by accreting companions.
Therefore cluster galaxies become gas poor and remain so and will evolve into
early-type galaxies.

\begin{figure} \centerline{\includegraphics[width=8cm]{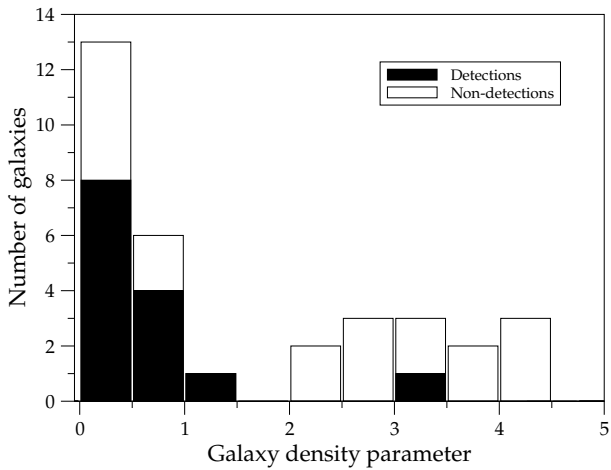}}
\caption{Distribution of the estimate of the local galaxy density as given in
the NCG for galaxies detected and not detected in \HI. Galaxies with galaxy
density above 2 are  member of the Virgo cluster} \label{fig:env} \end{figure}

Given the above, one might expected that the \HI\ properties of early-type
galaxies strongly depend on the local galaxy density, even more so than for
spirals.  Observational evidence that this is indeed the case comes from
studies based on the Arecibo Legacy Fast ALFA survey
\citep[ALFALFA][]{Giovanelli05}.  \cite{diSerego07} and \cite{Grossi09} have
used this survey to select early-type galaxies located in the Virgo cluster
and in low density environments respectively.  \cite{diSerego07} found a
detection rate of only 2\% for early-type galaxies that are member of the
Virgo cluster. On the other hand, the detection rate for the early-type
galaxies in low density environments \citep{Grossi09} is about ten times
higher (25\%). A similar contrast in \HI\ properties has been observed by
\citet{Serra09}.  Although a simple division into cluster and field galaxies
does not do justice to the wide range of \HI\ properties observed in field
galaxies (see below), our data, that moreover have a noise level about a
factor 4 better than the ALFALFA data, strongly confirm the large difference
in detection rates between galaxies in low and in high density environments.
To illustrate this, we have divided our sample into cluster and field
sub-samples and for the moment we ignore the wide range of the \HI\ properties
observed.  To quantify the environment of our sample galaxies, we have used
the estimates of their local galaxy densities given in the Nearby Galaxy
Catalog \citep[NGC,][]{Tully88}. This catalog gives, on a spatial grid of 0.5
Mpc, the density of galaxies brighter than --16 mag.  Applying this
information to our sample clearly separates it into low- and high-density
sub-samples (see Fig.\ \ref{fig:env}). The former sub-sample we will refer to
as the field sample and the latter corresponds to galaxies in the Virgo
cluster.

Figure \ref{fig:env} shows the distribution of the local galaxy density for
galaxies not detected in \HI\ and for galaxies where at least some \HI\ was
detected in or near the galaxy.  This figure shows that only one object (NGC
4262) out of 13 cluster galaxies (8\%) has been detected in \HI, while in, or
near, 14 of the 20 non-cluster galaxies (70\%) at least some \HI\ is detected,
albeit with a large variety of \HI\ properties ranging from small, off-centre
clouds to large, regular \HI\ discs. It is clear that, even with more sensitive
observations, it is much more likely to detect \HI\ in or near field galaxies
than in or near Virgo early-type galaxies.

An important observation is that, as for Virgo spirals, the detection rate of
{\sl molecular gas} in early-type galaxies shows a much weaker environmental
dependence \citep{Combes07,Young10}.  Since these CO observations mostly refer
to the central regions, this suggest that in early-type galaxies, gas removal,
as in cluster spirals, occurs predominantly from their outer regions. The
similarity of the effects of stripping suggest that gas removal is due to the
same mechanism.  The observation that the \HI\ detection rate of early-type
galaxies seems to be more affected than that of spirals is consistent
  with the fact that early-type galaxies form a more relaxed Virgo
  population than Virgo spirals \citep{Bingelli87}. The different dynamical
  properties of the two populations suggest that early-type galaxies have been
  member of the Virgo cluster for a longer period and therefore may have
  suffered more from the environmental effects that remove gas from galaxies.
  Another effect can be that, compared to spirals, the \HI\ in early-type
  galaxies is more often found in the outer regions of the galaxies so that it
  is more easily affected by interactions in the cluster. 

  Nevertheless, some of our detections in the field, e.g.\ NGC 3032 and NGC
  3489, have their \HI\ in an inner disc that could survive in the dense
  environment of Virgo. Still, such \HI\ discs are not seen in our
  observations of Virgo galaxies. This could be explained if such inner discs
  in field galaxies are the remnants of relatively recent accretions (see also
  next section). The flat slope of the faint-end of the \HI\ mass function
  observed for the Virgo cluster \citep{Kent10} indicates that the population
  of small gas-rich objects, i.e.\ those objects that appear to supply field
  early-type galaxies with fresh gas, is smaller in the Virgo cluster than in
  the field. Therefore, once a Virgo early-type galaxy loses its gas through
  environmental effects, there is much less chance that a new supply of gas is
  accreted and once a Virgo early-type galaxy is \HI\ poor, it will likely
  remain so, unlike galaxies in the field.  Even if a gas-rich object
  accreted, interactions with the high density environment will more likely
  remove such gas during the accretion while it is still only loosely bound to
  the galaxy. That accretions in clusters are less gas rich is also observed
  for shell galaxies by \citet{Hibbard03}.  The low gas accretion rate for
  cluster galaxies is also suggested by the observation that in several field
  galaxies we observe ongoing accretion, in stark contrast with the cluster
  galaxies, although the dense cluster environment implies a shorter time over
  which direct signs of interaction are visible. A lower detection rate of
  direct signs of ongoing interaction in clusters is also observed in other
  wave bands \citep[e.g.,][]{Tal09}. Both gas removal and the lack of new gas
  supplies drive the morphological evolution of galaxies in the Virgo cluster
  from late type to early type.

\subsection{Morphology of the \secHI\ in field early-type galaxies}
\label{sec:morphHi}
The results of the previous section show that there is a large difference in
overall \HI\ properties between cluster and field early-type galaxies.
However, as many earlier studies have shown, within the group of field
early-type galaxies there is a very large range in \HI\ properties and the
situation with respect to \HI\ in early-type galaxies is much more complex than
just a simple division into cluster and field.  Here we  discuss this in
some more detail.

Combining the results from \cite{Morganti06} and the present work, we have
deep \HI\ images for 20 field early-type galaxies selected from the SAURON
sample of 48. For 13 of these objects, we have detected \HI\ in or near them.
Figure \ref{fig:himaps}, together with Fig.\ 1 from \cite{Morganti06}, shows
that there is a very large range in properties of the \HI\ structures in
early-type galaxies, ranging from a single small cloud, to large, regular gas
discs.  Nevertheless, some overall trends can be seen and the morphology and
kinematics of the detected \HI\ structures can be divided into 3 broad
categories.

The first category (denoted C, for cloud) contains those objects where the
\HI\ is found in small clouds, where in some cases it is even not obvious
whether the \HI\ clouds are likely to be bound to the galaxy, or whether they
are, e.g., "free-floating" remnants of a past interaction or even a low
surface brightness companion. The galaxies that fall in this category are NGC
524, NGC 3384, NGC 3608, NGC 5982 and NGC 7332. The \HI\ masses involved are
small and are all less than a few times $10^7$ $M_\odot$. 

The second category (denoted A, for accretion) contains galaxies where the
\HI\ is found in unsettled structures that are clearly connected to a recent
or an ongoing gas-rich accretion. The galaxies that fall in this category are
NGC 1023, NGC 2768 and NGC 5198, while to some extent NGC 3489 could also fall
in this category. The \HI\ in NGC 1023  shows overall
rotation, but the kinematics shows many irregularities and the \HI\ is clearly
not settled.  The \HI\ masses involved range from several times $10^6$
$M_\odot$ to  over $10^9$ $M_\odot$.

The final category (denoted D, for disc/ring) refers to galaxies where most of
the \HI\ is found in a fairly regularly rotating disc or ring. This category
contains the galaxies NGC 2685, NGC 3032, NGC 3414, NGC 3489, NGC 4150 and NGC
4278, while also the only cluster galaxy detected (NGC 4262) falls in this
category.  In NGC 2685 and in NGC 4278 the \HI\ disc is large, i.e. extending
beyond the bright optical body, while in NGC 3032, NGC 3489 and in NGC 4150
the \HI\ forms a small, inner gas disc. In NGC 3414, the \HI\ appears to form
a polar ring or disc.  For galaxies with discs, the \HI\ masses involved range
from several times $10^6$ $M_\odot$ to over $10^9$ $M_\odot$.

The first conclusion that can be drawn is that in about half the galaxies
where \HI\ is detected, it is found in a disc-like structure. Based on the
first set of observations of the \sauron\ sample, \cite{Morganti06} had found
that gas discs appear to be common in early-type galaxies, while a similar
result was found for early-type galaxies detected in the HIPASS survey
\citep{Oosterloo07}. The extended set of observations of the \sauron\ sample
clearly confirms this.

The second conclusion is that accretion of gas is common in early-type field
galaxies. For all galaxies detected the \HI\ data reveal that accretion is
on-going, or has occurred in the recent past. Galaxies with a gas disc also
show clear signs of accretion: in NGC 3489, a faint, extended gas tail is
detected which connects to the inner gas disc. Similarly, near NGC 4150 a small
\HI\ cloud is detected, while even the outer regions of the large, and
presumably older, \HI\ disc in NGC 4278 are connected to two large gas tails.
The disc in NGC 2685 is heavily warped while the disc in NGC 3414 is polar,
hence it is likely that also in these galaxies the gas has been accreted. The
gas disc in NGC 3032 is counter-rotating to the stars. The conclusion is that
gas discs in early-type galaxies form through accretion and that this is an
ongoing process. Some of the variation we see in \HI\ properties in our sample
may be reflecting different stages of such accretion events. For example, NGC
2768 is accreting gas and a small inner polar disc is forming
\citep{Crocker08}. It is quite possible that this galaxy will evolve into a
system similar to NGC 3414. The accretion and inner discs in NGC 3489 and NGC
4150 appear to have a similar history, where NGC 4150 is probably at a
slightly more evolved stage. NGC 3032 may be at an even more evolved stage.
Another example is NGC 1023 where a large and fairly massive \HI\ structure is
observed that shows an overall rotation pattern, but that clearly has not
settled into a disc. This system may evolve into a galaxy with a large,
regular gas disc, similar to the one seen in NGC 4278.

In a few galaxies, the amount of \HI\ detected is above $10^9$ $M_\odot$,
i.e.\ similar to the amount of \HI\ in the Milky Way.  This suggests that in
those cases the object accreted must have been fairly massive. Moreover, the
large extent and regular disc kinematics indicate that some must have formed
several Gyr ago.  However, in most galaxies the \HI\ masses involved are
smaller and correspond to that of galaxies like the Magellanic Clouds or
smaller. Assuming $10^9$ yr for the timescale of a typical accretion
\citep[e.g.][]{Sancisi08,Tal09}, the detection rate and observed \HI\ masses
imply that the accretion rate for cold gas is smaller than 0.1 $M_\odot$
yr$^{-1}$ for most field early-type galaxies.  This suggests that, even
allowing for the fact that the \HI\ is only a fraction of the mass accreted,
currently early-type galaxies grow only by a modest amount through accretion.
This has to be the case because, although most galaxies are small in size,
most of the mass in galaxies is already in large galaxies \citep{Renzini06}.
Therefore there is no large enough reservoir of small galaxies available for
large galaxies to grow substantially by accretion of companions. We note that
only one galaxy (NGC 2685) shows clear optical peculiarities that can be
associated with accretion (in this case polar dust lanes). This underlines
that, although accretion often occurs, the mass of the accreted object is, in
most cases, small compared to that of the host galaxy and the effects on the
host galaxy usually are at most subtle.  This is in line with optical imaging
studies of other samples of early-type galaxies
\cite[e.g.][]{Schweizer92,Dokkum05,Tal09} which have shown that most
early-type galaxies in the field and in small groups show signs of small
accretion events. The new aspect from our results is that small amounts of gas
are involved in this continuing assembly of field early-type galaxies.

In a recent review, \cite{Sancisi08} concluded that for at least 25\% of field
spiral galaxies there is direct evidence that a small gas-rich companion or
gas cloud is accreting, or has been accreted in the recent past. Also for many
of these objects the optical image does not show clear signs of interaction
and the accretion is only visible in \HI\ observations.  It appears that, as
far as the character of accretion is concerned, there is not much difference
between field spiral galaxies and field early-type galaxies and to some
extent, the \HI\ properties of early-type galaxies bear a resemblance with
those of the outer regions of spiral galaxies. \cite{Sancisi08} estimate that,
for field spiral galaxies, the accretion rate for cold gas is about 0.1 to 0.2
$M_\odot$ yr$^{-1}$, somewhat higher than we estimate for our early-type
galaxies.

\subsection{Dynamical structure of the host galaxies}

\label{sec:envDynamix}

Several theoretical studies have shown that the dynamical structure of a
merger remnant critically depends on the amount of gas, and hence dissipation,
present in (one of) the progenitors
\citep[e.g.][]{Bender92,Jesseit05,Naab06,Hopkins09,Jesseit09}. Moreover, a
systematic change of the importance of dissipative effects as function of mass
may explain the different dynamical properties of high-mass and low-mass
early-type galaxies \citep{Davies83}.

In \cite{Morganti06} we had found that the \HI\ detections are uniformly
spread through the $(\epsilon, V/\sigma)$ diagram and we concluded, although
the sample used was small, that if fast and slow rotators represent the relics
of different formation paths, this did not appear to be reflected in the {\sl
  current} characteristics of the \HI.  The discussion of the previous section
showed that, except in a few cases, the \HI\ currently detected is mainly due
to recent small accretions. Because the dynamical structure of a galaxy is the
result of the evolution over a Hubble time, a clear observable link with these
recently accreted small amounts of gas may not be expected.

For our extended dataset there is, similar to the result of \citet{Morganti06},
not much evidence that, for most galaxies, the current \HI\ content is
connected to the dynamical characteristics of the galaxy.  Different from
\citet{Morganti06}, we use the parameter $\lambda_{\rm R}$, introduced by
\citet{Emsellem07} to describe the importance of rotation. This parameter
involves luminosity weighted averages over the full two-dimensional stellar
kinematic field as a proxy to quantify the observed projected stellar angular
momentum per unit mass. It can have values between 0 and 1, and apart from
projection effects, higher values of $\lambda_{\rm R}$ suggest that rotation
is more important for the dynamics of the galaxy.

\begin{figure} 
\centerline{\includegraphics[angle=0,width=8cm]{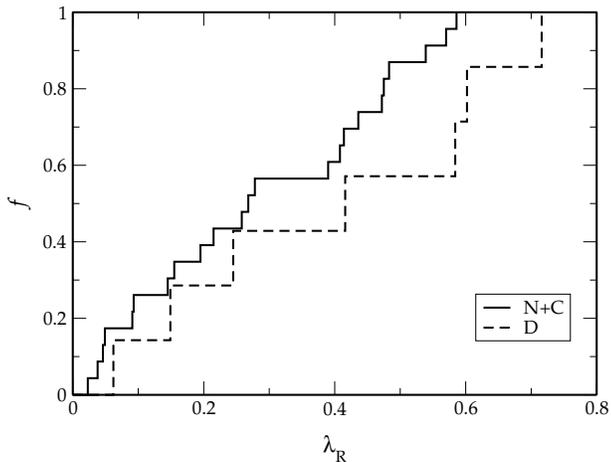}} 
\caption{Cumulative distribution of $\lambda_{\rm R}$ for galaxies classified
  as having an \HI\ disc (class D) compared with that for galaxies undetected
  in \HI\ or where only a small \HI\ cloud was detected (classes N and C).}
\label{fig:rotator} 
\end{figure}

In Fig.\ \ref{fig:rotator} we show the cumulative distributions of
$\lambda_{\rm R}$ for the group of galaxies that one could see as fast {\sl
  gas} rotators (i.e.\ class D) and for the group of galaxies that are \HI\
non- or slow gas rotators (classes N and C). No clear dichotomy emerges. For
the discy \HI\ detections, the distribution of $\lambda_{\rm R}$ appears to go
to higher values, which would suggest that for some galaxies with an \HI\ disc
rotation is also important for the stellar component. On the other hand, there
is good overlap for smaller values of $\lambda_{\rm R}$ suggesting that the
presence of an \HI\ disc is not a good discriminator. This is also suggested
by, for example, the fact of the sixteen fast rotating field galaxies, only
six have the \HI\ distributed in a disk.

\begin{figure} 
  \centerline{\includegraphics[angle=0,width=8cm]{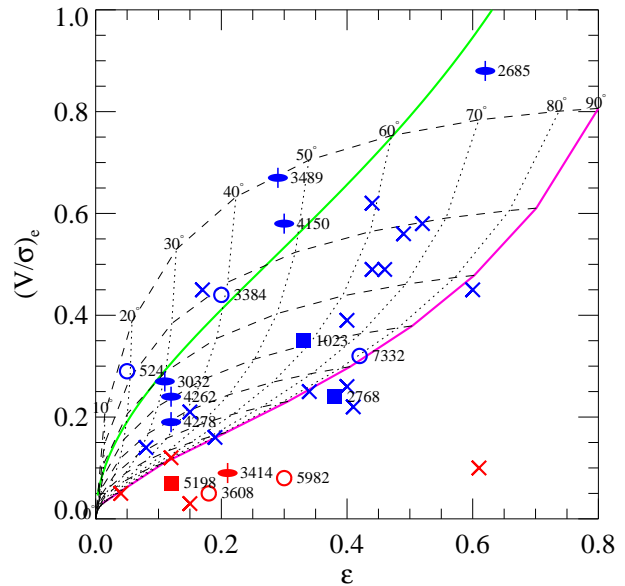}}
  \caption{$(V/\sigma,\varepsilon)$ diagram for galaxies in our HI sample.
    Crosses are non detections, Ellipses with axis are are HI disks, Empty
    circles are isolated HI clouds and filled squares are HI accretion.
    Red/blue colours indicate slow/fast rotators. The NGC numbers of the HI
    detected galaxies are indicated. The magenta line corresponds to the
    relation $\delta=0.7\varepsilon_{\rm intr}$ between anisotropy and
    intrinsic ellipticity. The dotted lines indicate how the relation is
    transformed when the inclination (indicated at the top) is decreased. The
    values are taken from \citep{Cappellari07}, where a detailed explanation
    of the diagram is given.  }
\label{fig:VS} 
\end{figure}

The \sauron\ galaxies were selected to sample uniformly the {\em projected}
axial ratio and the E/S0 morphology. Both quantities are related to the galaxy
inclination as is $\lambda_R$. For this reason the distribution of $\lambda_R$
is related in a complex way to the selection criteria and it is difficult to
interpret quantitatively. An alternative way to look at the relation between
dynamics and \HI\ morphology, while including possible inclination effects, is
by using the $(V/\sigma,\varepsilon)$ diagram \citep{Binney05}. The study of
the distribution of the \sauron\ galaxies on the diagram was discussed in
\citet{Cappellari07}. It was shown that fast-rotators ETGs tend to lie in a
restricted region of that diagram, defined at the lower boundary by a linear
trend between intrinsic flattening and orbital anisotropy for edge-on systems
(the magenta line in Fig.\ \ref{fig:VS}). Lowering the inclination moves
galaxies to the left of that line on the $(V/\sigma,\varepsilon)$ diagram as
illustrated in Fig.\ \ref{fig:VS} (for a detailed explanation see
\citet{Cappellari07}).  By plotting our different \HI\ morphologies on the
diagram we conclude that: (i) The galaxies with the fastest intrinsic
(edge-on) $V/ \sigma$ in our sample all happen to posses \HI\ disks. To these we
should add NGC~2974, which has an intrinsic $V/\sigma$ larger than NGC~3489
and also posses an \HI\ disk \citep{Weijmans08}. Within our limited number
statistics this would suggests that an accretion/merger involving a large
amount of gas is required to produce the galaxies most dominated by rotation;
(ii) However the reverse is not true as \HI\ disks can be present also at
intermediate $V/\sigma$, which implies that \HI\ disks do not necessarily
produce a rotation-dominated object. A regular disk is seen in fact even in
the slow rotators NGC~3414. Larger samples are needed to conclusively
understand the link between \HI\ and galaxy dynamics, but this work illustrates
that there is not a simple connection.

\subsection{\secHI\ and molecular gas} \label{sec:molecular}

Atomic hydrogen is not the only tracer of the cold ISM in galaxies. A CO
survey of the \sauron\ sample \citep{Combes07} has shown that a significant
fraction of the \sauron\ galaxies contains molecular gas. It is therefore
interesting to compare the \HI\ properties of the \sauron\ galaxies with those
of the CO. More recent observations (see Tab.\ \ref{tab:tableCO}) have
somewhat modified the list of CO detections from \citet{Combes07}.  The
earlier CO detections in NGC~4278 and NGC~7457 have not been confirmed, while
CO has now been detected in NGC~524. Of the sample we discuss in this paper,
six galaxies are detected both in CO and in \HI\ (NGC 524, NGC 2685, NGC 2768,
NGC 3032, NGC 3489 and NGC 4150), out of a total of nine CO detections and
fifteen \HI\ detections. Interestingly, all these galaxies show small amounts
of star formation \citep[][see also Sect. 4.6]{Shapiro10}. In five of the CO
detections \HI\ is also seen in the central region of the galaxy (NGC~2685,
NGC~2768, NGC~3032, NGC~3489, NGC~4150) and the CO and the \HI\ show very
similar kinematics, indicating that the same component is detected both in CO
and \HI. The exception is NGC 524 where no atomic counterpart is detected of
the central molecular disc. Similarly, in the Virgo galaxies NGC~4459,
NGC~4477 and NGC~4550 CO was found, but no \HI.

\begin{figure} \includegraphics[angle=270,width=8cm]{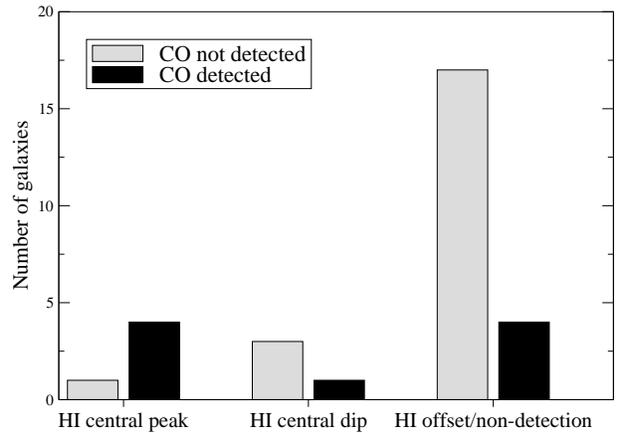}
\caption{CO detection histogram as a function of central \HI\ morphology.}
\label{fig:molecular1} \end{figure}

For a proper comparison of the \HI\ and the H$_2$ properties, it is important
to keep in mind that the \HI\ observations typically probe a region of
several tens of kpc in size, i.e.\ an entire galaxy and its immediate
environment, while the CO observations refer only to the  central region
of a galaxy on kpc scales. A comparison based on total \HI\ content and a
central CO measurement is not necessarily meaningful. The spatial resolution
of our WSRT observations is well matched to the field-of-view of the IRAM 30-m
dish used by \citet{Combes07}. We therefore compare the CO measurements with
the \HI\ detected in the single WSRT beam centred on the galaxy while taking
into account the central morphology of the \HI\ of our galaxies. We divided
the central \HI\ morphology in three categories: no central \HI, central dip in
the \HI\ (within a large-scale \HI\ structure), or central \HI\ peak. The
category for each galaxy is listed in Table \ref{tab:tableCO} and the relation
with the CO properties is given in Fig.\ \ref{fig:molecular1}. This figure
clearly shows that a statistical relation exists between the presence of \HI\
and CO in the central regions. Galaxies with centrally peaked \HI\ are 
more likely to have CO than the other two types. Of the six galaxies with a
centrally peaked \HI\ distribution, five have a central CO component. For
comparison, if one would use the total \HI\ content, the statistics clearly
gets diluted: of all fifteen \HI\ detections, only six are detected in CO.  We
conclude that centrally peaked distributions of \HI\ tend to harbour
corresponding central molecular gas distributions. The only exception to this
rule seems to be NGC 3414. However, the kinematics of the \HI\ in this galaxy
suggests that this \HI\ likely forms an edge-on polar ring and that the
observed central \HI\ peak may be due to projection.

\begin{table} 
\tabcolsep=3pt 
\begin{center}
\begin{tabular}{cccccc} 
\hline\hline NGC     &\HI\ type & $M_{\rm H_{2}}$   &$M_{\rm H_{2}}$/$M_{\rm HI}$ & CO references \\ 
&  & $M_\odot$         & central   &      \\ 
(1)     &(2) &  (3)       & (4)      & (5)  \\ 
\hline 
~524  & none   & $1.6 \times 10^{8}$            & \llap{$>$}25.7  & a \\ 
1023  & dip    & \llap{$<$}$4.0 \times 10^{6}$  & \llap{$<$}0.6   & b \\ 
2685  & dip    & $2.1 \times 10^{7}$            &        12.2     & c \\ 
2768  & peak   & $6.8 \times 10^{7}$            &        8.5      & d \\ 
3032  & peak   & $5.0 \times 10^{8}$            &        11.1     & e \\ 
3414  & peak   & \llap{$<$}$9.0 \times 10^{6}$  & \llap{$<$}0.29  & b \\ 
3489  & peak   & $1.2 \times 10^{7}$            &        8.8      & b \\ 
4150  & peak   & $5.5 \times 10^{7}$            &        21.0     & e \\ 
4262  & dip    & \llap{$<$}$1.1 \times 10^{7}$  & \llap{$<$}2.32  & b \\
4278  & dip    & \llap{$<$}$6.4 \times 10^{6}$  & \llap{$<$}1.32  & a \\ 
4459  & none   & $1.6 \times 10^{8}$            & \llap{$>$}40.6  & e \\ 
4477  & none   & $2.4 \times 10^{7}$            & \llap{$>$}6.0   & b \\ 
4550  & none   & $7.2 \times 10^{6}$            & \llap{$>$}1.9   & f \\ 
\hline 
\end{tabular} 
\caption{Molecular mass to \HI\ mass ratios for the galaxies with CO detections
and/or central \HI. (1) Galaxy identifier. (2) Classification of the central \HI. (3)
Total molecular gas mass. (4) Ratio of total molecular gas mass to the \HI\ mass
observed in the central interferometric beam. (5) References:
(a) Crocker et al.\ unpublished; (b) \citet{Combes07}; (c)
\citet{Schinnerer02}; (d) \citet{Crocker08}; (e) \citet{Young08} (f)
Crocker et al.\ (in preparation).
\label{tab:tableCO}} 
\end{center}
\end{table}

A few more things can be learned from this comparison. In the five galaxies
where both \HI\ and CO are detected in the central regions, interferometric CO
observations \citep{Crocker08,Young10,Crocker10} show that the kinematics and
the morphology of both components are clearly connected and the same physical
structure is detected. A very nice example is the inner gas disc of NGC 3032
for which the extent and kinematics as seen in CO and in \HI\ match exactly. A
very interesting aspect is that the combination of the CO and \HI\
observations of NGC 2768, NGC 3032, NGC 3489 and NGC 4150 gives clear evidence
that these inner gas structures form by the accretion of small-gas-rich
objects. The inner gas discs that are seen in NGC 2768, NGC 3489 and NGC 4150
connect to large, faint \HI\ plumes that are seen at larger radii, clearly
showing that these gas structures formed from accreted gas. The fact that the
inner gas disc in NGC 3032 is counter-rotating to the stars also shows an
external origin. Our results also show that the cold ISM of the inner gas
discs detected both in \HI\ and CO is mainly in the form of molecular gas,
with the molecular gas mass being about 10 times higher than that of the
atomic gas. Interestingly, this mass ratio is very similar to that seen in the
centres of nearby spiral galaxies \citep[e.g.][]{Leroy08}, despite the very
different state of the ISM in these two types of galaxies.

\begin{figure*} 
\centerline{
\includegraphics[angle=0,width=8cm]{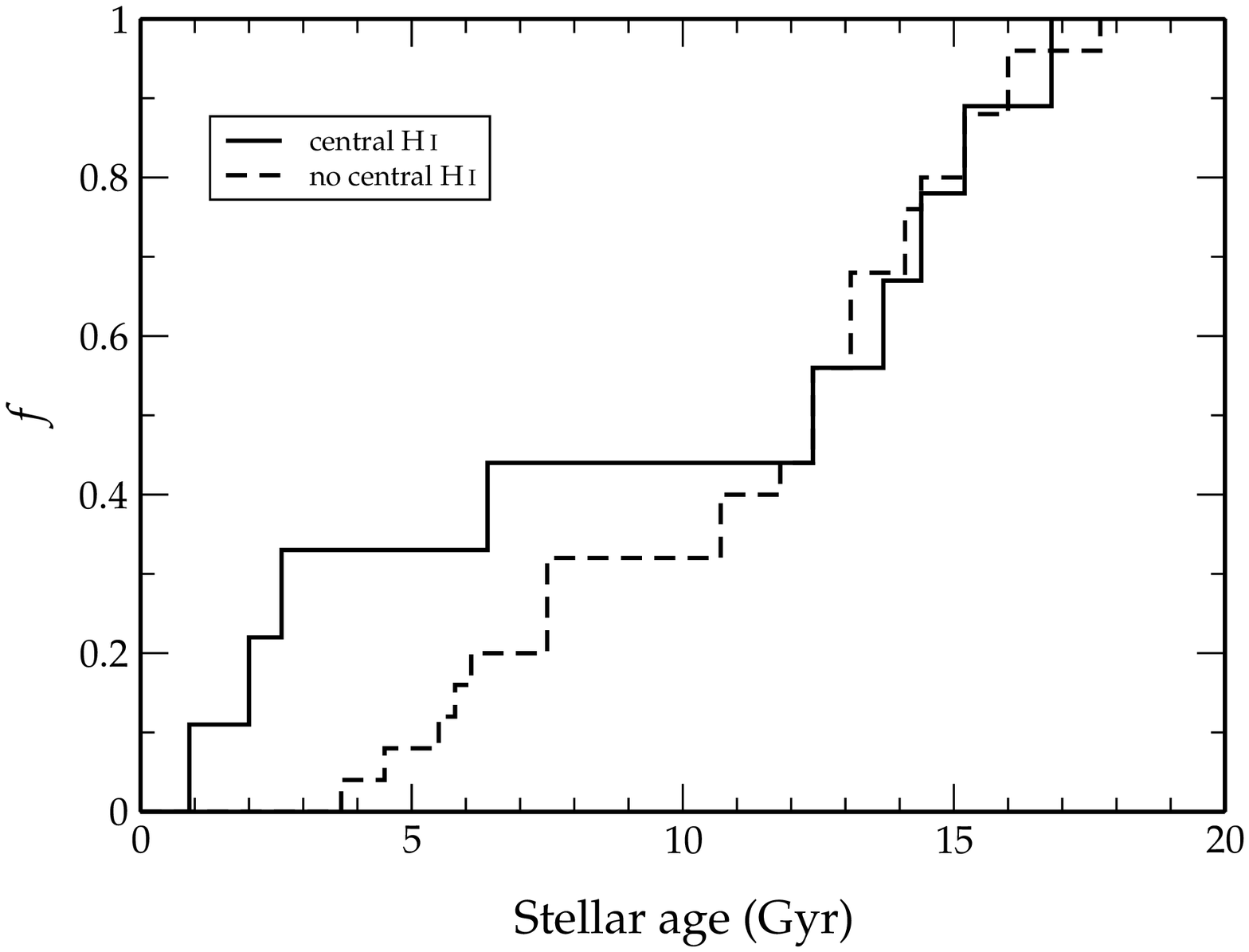}\hss
\includegraphics[angle=0,width=8cm]{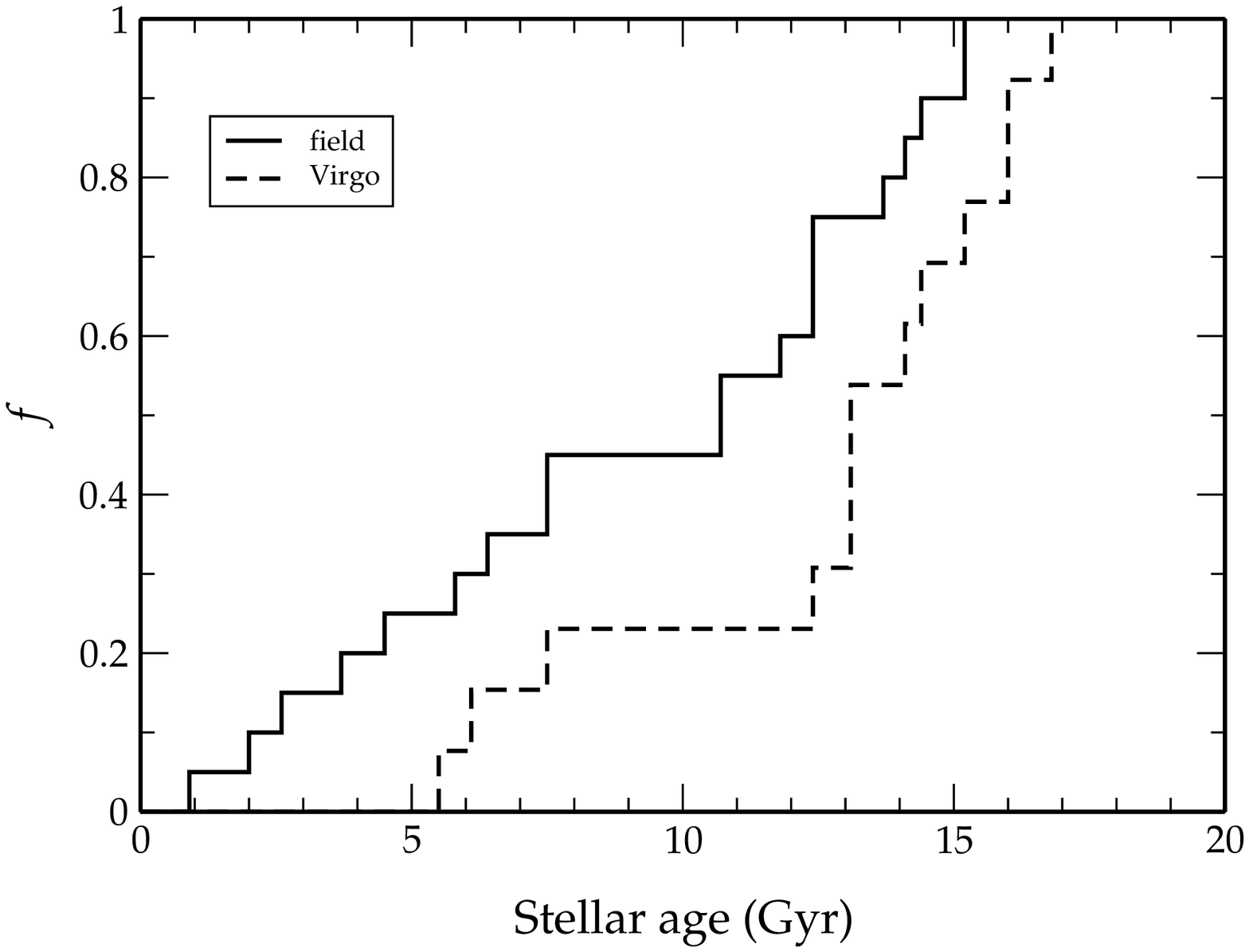}
}
\caption{Left: cumulative distributions of stellar ages inside 1 $R_e$ from
  single-population models from \citet{Kuntschner10} where the sample
  is split in those galaxies with and those without \HI\ detected in their
  central region. Right: Same distributions, but now the sample is split in
  field and Virgo galaxies }
\label{fig:ageCum} 
\end{figure*}

\subsection{\secHI\ and ionised gas}

With the new \HI\ observations presented in this paper, we further confirm the
result of \citet{Morganti06} that galaxies with regular \HI\ discs tend
to have strong, extended emission from ionised gas that has the same
kinematics as the \HI\ while galaxies with unsettled \HI\ structures have less
ionised gas.  For the galaxies for which we  present data for the first
time, regular discs of ionised gas are detected with \sauron\ in NGC~3032,
NGC~3489 and NGC~4262 \citep{Sarzi06}, exactly the three galaxies in the new
observations in which regular \HI\ discs are found. In the remaining \HI\
detections, where the \HI\ is found in small clouds offset from the centre
(NGC~524, NGC~3384 and NGC 3608), only small amounts of ionised gas are
detected. We therefore reiterate that, in early-type galaxies, a regular \HI\
disc always has an ionised counterpart.

\subsection{Neutral hydrogen and stellar population}
\label{sec:stellarPop}

Several earlier studies \citep[e.g.][]{Serra06,Morganti06} have concluded that
the relation between \HI\ content and stellar population for field early-type
galaxies is complex.  Our data show that, on the scale of the spatial
resolution of our data, the column densities of the \HI\ are below the
threshold above which star formation is generally occurring. However, the
highest column densities observed are not very much below this threshold, so
given the low spatial resolution of our data, some star formation could occur
in smaller regions. Low-level star formation (typically below levels of 0.1
$M_\odot$ yr$^{-1}$) is indeed observed in some of the gas-rich \sauron\
galaxies \citep[][see also below]{Temi09,Jeong09,Shapiro10}.

First we discuss to what extent the \HI\ properties are connected to the
ongoing star formation. As expected, some relation seems to hold. This is best
illustrated by comparing our data with those from \citet{Shapiro10}. Based on
{\it Spitzer} data, \citet{Shapiro10}, similarly to \citet{Temi09}, observe
small amounts of star formation in a subset of the \sauron\ galaxies. The star
formation they observe only occurs in galaxies characterised by fast rotation.
The morphology of the star forming regions is that of a thin disc or ring.
Moreover, they were able to distinguish two modes of star formation. In one
mode, star formation is a diffuse process and corresponds to widespread young
stellar populations and high specific molecular gas content. \citet{Shapiro10}
associate this star formation with small accretion events.  In the second
mode, the star formation is occurring more centrally concentrated, while
outside the region of star formation only old stellar populations are present.
Smaller amounts of molecular gas are connected to these central star formation
events.  \citet{Shapiro10} speculate that in at least some of these objects,
the star are forming from gas resulting from internal mass loss.

It is interesting to see that in 4 of the 5 galaxies classified by
\citet{Shapiro10} to have widespread star formation and that are observed by
us, both CO {\sl and} \HI\ is detected in the central regions (Table
\ref{tab:tableCO}).  In contrast, in none of the galaxies with centrally
concentrated star formation \HI\ is detected in the centre. This confirms the
suggestion by \citet{Shapiro10} that widespread star formation in early-type
galaxies is connected with higher gas content. We do note that for some
early-type galaxies {\it GALEX} data show that this relation between gas
content and star formation also exists at {\sl large} radius.  Examples are
NGC 404 \citep{Thilker10}, NGC 2974 \citep{Weijmans08} and ESO 381--47
\citep{Donovan09}.  Almost all galaxies with widespread star formation and
observed by us, have \HI\ discs (NGC 2685, NGC 3032, NGC 3489, NGC 4150),
quite distinct from galaxies with central star formation (no detection of NGC
4459 and NGC 4477, while only a small, offset \HI\ cloud is detected in NGC
524).  The only exception is NGC 4550, a galaxy in the Virgo cluster with
widespread star formation and not detected in \HI. This is a very unusual
galaxy with two counter-rotating stellar discs. We also note that in all four
galaxies with widespread star formation and detected by us, the \HI\ is likely
to have been accreted (see Sect.\ \ref{sec:morphHi}).  This confirms the
conclusion of \citet{Shapiro10} that widespread star formation is associated
with accretion.

\begin{figure}
\includegraphics[angle=0,width=8cm]{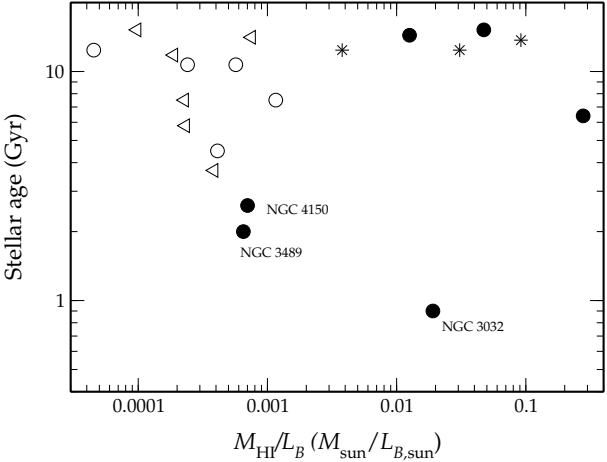}
\caption{Stellar age inside 1 $R_{\rm eff}$ plotted as function of relative
\HI\ content. The different symbols refer to \HI\ morphology: 
filled circles: regular disc/ring (class D); stars: unsettled \HI\ structures
(class A); open circles: small or offset \HI\ clouds (class C); triangles:
upper limits }
\label{fig:ageScatter} \end{figure}

The above results show that some star formation can occur in the gas
reservoirs of early-type galaxies, although not all gas-rich galaxies show
star formation. Overall, the connection with the properties of the stellar
population is poor.  Some galaxies indeed behave according to the expectation
that the presence of a relatively large amount of \HI\ indeed is connected to
the properties of the stellar population, but the rule seems to be that for
every rule there are exceptions.  For example, NGC 1023, NGC 3414 and NGC 4278
have extensive reservoirs of neutral hydrogen, but do not show any evidence
for the presence of a young stellar subpopulation.  To characterise the
overall stellar populations, we have used the single-population-equivalent
stellar ages inside 1 $R_{\rm eff}$ as derived for the \sauron\ galaxies by
\cite{Kuntschner10}. One way to investigate the connection between
\HI\ and stellar population is to compare these stellar ages of galaxies with
\HI\ to those of galaxies without central \HI. In Fig.\ \ref{fig:ageCum} we
give the cumulative distributions of this stellar age for galaxies where \HI\
is detected in the central regions and for those with no central \HI. This
figure suggests that some difference may exist between the two groups of
galaxies, with some gas-rich galaxies having younger stellar ages.  However,
the difference between the distributions is mainly caused by only a few
gas-rich galaxies that have quite young stellar ages (i.e.\ NGC 3032, NGC 3489
and NGC 4150 which are known to have ongoing star formation). Many galaxies
with central \HI\ have similar stellar populations as gas-free galaxies. Some
galaxies with fairly young or intermediate ages are in fact gas free (e.g.\
NGC 3377 and NGC 7457) and some gas-rich galaxies have large stellar ages. A
similar trend is seen when the total \HI\ content is used instead of the
central one.  This is further illustrated in Fig.\ \ref{fig:ageScatter} where
we plot the relative global gas content versus stellar age for the different
\HI\ morphologies.  No clear overall trend is visible in this figure, except
(again) that the three youngest galaxies all have a central \HI\ disc that is
also detected in CO.  The at most weak connection between \HI\ and stellar age
is to a large extent similar to that seen for the ionised gas.
\citet{Emsellem07} show that those \sauron\ galaxies with relatively much
ionised gas also have young stellar ages, but on the other hand several
galaxies with young or intermediate age populations are free of ionised gas.

The results discussed above clearly show that the relation between \HI\ and
stellar populations is complex. This is not unexpected, since the stellar
population is the result of the evolution of a galaxy over its entire life,
while the current \HI\ content only reflects the present state. In Sect.\ 4.2
we showed that some amounts of gas are accreted by field early-type galaxies
at irregular intervals. Often the amounts of gas involved are small and, even
if entirely converted into stars, will only leave subtle signatures in the
stellar population that are not always easy to detect \citep[see also
e.g.][]{Serra10}.  Moreover, the efficiency with which accreted gas is turned
into stars depends on many factors. For example, accretions characterised by
loss of gas angular momentum (e.g., retrograde encounters) result in efficient
gas infall that triggers central star formation. This may be observed as a
young population in a relatively gas-rich galaxy, but after a while, the gas
reservoir will be exhausted and the remnant is observed as \HI-poor and
centrally rejuvenated \citep{Serra06}. On the other hand, interactions in
which gas retains its angular momentum (e.g., prograde encounters) result in
large \HI\ tidal tails that can later be re-accreted to form large \HI\ discs.
Because of their large extent, the column densities in these discs is low and
at most some star formation will occur in localised regions at large radius
\citep[see e.g.][]{Oosterloo07}.  Recent work also suggest that bulges can
have a stabilising effect on discs, preventing star formation, even when
significant amounts of gas is present \citep{Martig09}.

It is instructive to consider the effect of the environment. The results
discussed so far show that the relation between {\sl current} gas content and
stellar population is complex. However, by comparing the Virgo early-type
galaxies with those in the field, one may get an idea of the effects of gas
accretion over longer periods of time. Our data indicate that for field
early-type galaxies, gas accretion does play a significant role in determining
the stellar population. This can be seen by comparing the cumulative
distributions of the stellar ages of Virgo and field galaxies (Fig.\
\ref{fig:ageCum}). These distributions show that there is a trend of field
early-type galaxies, as a population, having younger stellar ages than Virgo
early-type galaxies.  This difference likely reflects the different long-term
accretion history of the two groups of galaxies. Field early-type galaxies
regularly accrete small amounts of gas from their environment and, over time,
this leaves an observable signature in the stellar population.  In contrast,
galaxies in Virgo, being in a gas poor environment, grow much less by
accretion of gas-rich companions and even if some gas is accreted, it is
removed on a short timescale by the cluster environment.  The stellar
population of cluster early-type galaxies is much less rejuvenated during
their evolution compared to field early-type galaxies.

\subsection{\secHI\ and the radio continuum}

\label{sec:comparison_agn}

As described in Sec.\ 2, our observations have also allowed to extract images
of the radio continuum emission. As detailed below, these images are in many
cases much deeper than those available so far. This resulted in a significant
number of new detections of the radio continuum associated with our target
galaxies. We detect radio continuum in 13 of the 20 field galaxies and in 5 of
the 13 cluster galaxies.  With the exception of a few well-known objects with
strong radio continuum (e.g.\ NGC~4278, NGC~4374/M84), the majority of the
detected sources have a radio flux of at most a few mJy. In all detected
sources, the radio continuum comes from the central region of the galaxy.

\begin{figure}
\includegraphics[angle=-90,width=8cm]{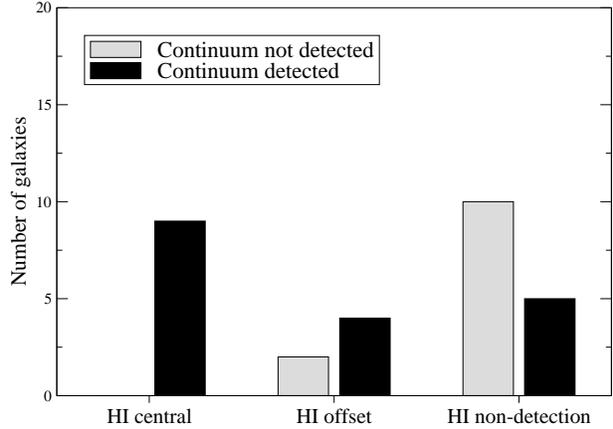}
\caption{Histogram of the distribution of radio continuum detection as function
of \HI\ detection and morphology. 
}
\label{fig:agn}
\end{figure}

The interesting result is that there appears to be a trend between detection
of radio continuum and detection of \HI\ in or around the galaxy. Figure
\ref{fig:agn} shows the distribution of radio continuum detections as function
of \HI\ presence and \HI\ morphology. The histogram shows that the radio
continuum detection rate is higher for objects where also \HI\ is detected,
with a suggestion of an additional trend that galaxies with \HI\ but not in
their central regions are less likely to be detected in radio continuum than
galaxies with \HI\ in the centre, but more likely than galaxies with no \HI\
at all. This suggests that the cold gas somehow contributes in feeding the
processes that produce the radio continuum emission in some galaxies. It is
well known that both star formation and AGN activity can contribute to the
radio continuum emission from early-type galaxies \citep{Wrobel91}. Here we
address the question is to what extent the observed trend with HI properties
is connected to radio emission from star formation or from radio-loud AGN.

To study this question, we estimated the radio continuum flux expected to be
observed from the amounts of star formation seen by \citet{Shapiro10} and
compared these with the observed fluxes.  Of the 8 star forming galaxies
overlapping between \citet{Shapiro10} and us, 7 are detected in radio
continuum. We used the relation given by \citet{Bell03} to convert star
formation rates to radio luminosities. For NGC 2685, NGC 3032, NGC 3489, NGC
4150 and NGC 4459, the observed radio continuum flux matches, within the
errors, what is expected from the observed star formation rates and we
conclude that the observed radio continuum is associated with the star
formation.  NGC 524 is known to have a faint radio AGN \citep{Nagar05}, and
when correcting the observed radio flux for this AGN, also for this galaxy the
observed radio continuum matches that what is expected based on the observed
star formation rate.  Our upper limit to the continuum flux of NGC 4550 is
consistent with the observed star formation in this galaxy.  Because
\citet{Shapiro10} derive the star formation rates from FIR data, the good
match between predicted and observed radio continuum in these galaxies implies
that they follow the well-known radio-FIR correlation. Only in one star
forming galaxy (NGC 4477) is the observed radio flux much larger than expected
(by about a factor 7), which most likely means that this galaxy harbours a
radio loud AGN.

In the previous section, we showed that galaxies with \HI\ in their central
regions are more likely to have star formation, while here we find that in
many of these star forming galaxies, the radio continuum detected is due to
this star formation.  For about half the galaxies represented in the first bin
of Fig.\ \ref{fig:agn} this is the case.  On the other hand, for both other
\HI\ classes shown in this figure, only one of the objects is forming stars.
These results show that the trend seen in Fig.\ \ref{fig:agn} is at least
partly due to enhanced star formation in galaxies with central \HI.

The spatial resolution of our observations is relatively low (corresponding to
a linear scale of the order of a kpc for the most distant galaxies). This
makes it difficult to decide, based on the radio data alone, whether the
observed radio continuum is due to a radio-loud AGN.  We therefore have
searched the literature for higher resolution radio continuum data.

One possibility is to look into the FIRST survey \citep[Faint Images of the
Radio Sky at Twenty-Centimeters][]{Becker95}. FIRST data have  better spatial
resolution ($\sim 5^{\prime\prime}$), but  higher noise level (about 1 mJy beam$^{-1}$)
compared to our WSRT observations.  A search for detections in this survey has
been also done by \citet{Sarzi10}.  Of the 9 field galaxies detected by us at
the level of a few mJy (i.e.\ such that the sensitivity of FIRST would allow
to detect these sources) 7 are detected by FIRST. In only one of these
detections - NGC~3032 - the radio emission is resolved by FIRST, suggesting
the presence of star formation, consistent the results presented here on this
galaxy.

More instructive is, however, to consider the work of \citet[][and references
therein]{Nagar05}. Their work focuses on detecting radio nuclei with high
brightness temperature ($> 10^7$ K) and/or jet-like structures as unambiguous
indications for the presence of an AGN.  They concentrated on low-luminosity
active galactic nuclei (LLAGN) and AGN selected from the Palomar Spectroscopic
sample of northern galaxies that they have studied at high frequency (15 G~Hz)
with the VLA (0.15$^{\prime\prime}$ resolution) and with VLBI at intermediate
and high radio frequencies. At these high resolutions (corresponding to a
linear size of a few tens of pc), they detect radio emission in almost 50 \%
of the sources, suggesting a high incidence of radio cores.

Interestingly, their sample includes a fair number of our objects, although,
unfortunately, the sensitivity of their observations is not as good as of our
WSRT observations (flux limit of 1-1.5 mJy for the VLA observations and 2.7
mJy for the VLBI observations).  Of the galaxies that are in common, for 11 is
our WSRT flux above their detection limit and {\em all} are detected in the
high-resolution observations of Nagar et al., suggesting the presence of a
compact (core?) structure in these galaxies.  One further object (NGC 5198) is
not included in the Nagar et al.\ list but is classified as AGN by
\citet{Sarzi10} and also this object is detected in radio (both by WSRT and by
FIRST). Almost all these objects are observed not to have star formation
\citep[e.g.][]{Shapiro10}. We conclude that many of our early-type galaxies
harbour a radio-load AGN. Interestingly, these AGN are evenly distributed over
the three \HI\ classes used in Fig.\ \ref{fig:agn} suggesting that AGN
activity is not connected to whether there is cold gas in the central region
of an early-type galaxy or not. The observed relation between central gas
content and radio continuum seems therefore to be due to a higher probability
for radio emission from star formation if a galaxy has a relatively large
amount of gas in the central regions.

\section{Conclusions}

\label{sec:conclusions}

We have presented new, deep Westerbork Synthesis Radio Telescope observations
of the neutral hydrogen in 22 nearby early-type galaxies selected from a
representative sample of early-type galaxies studied earlier at optical
wavelengths with the \sauron\ integral-field spectrograph. Combined with our
earlier observations, this resulted in deep \HI\ data on 33 nearby early-type
galaxies. This is the largest homogeneous dataset of \HI\ imaging data
available for this class of objects.

In contrast to our earlier study \citep{Morganti06}, this sample both covers
field environments and the Virgo cluster.  We find that the \HI\ properties 
strongly depend on environment. For detection limits of a
few times $10^6$ $M_\odot$, we detect \HI\ in about 2/3 of the field galaxies,
while for Virgo  galaxies the detection rate is $<$10\%. This is
consistent with earlier work \citep{diSerego07,Grossi09}.  We confirm earlier
results \citep{Morganti06,Oosterloo07} that in about half the early-type
galaxies with \HI, this atomic gas is in a regularly rotating \HI\ disc or
ring.  In many  objects, unsettled \HI\ structures are detected,
such as tails and clouds, sometimes connecting to a regular \HI\ disc. This
suggests that the gas discs form through accretion. We conclude that accretion
of \HI\ commonly occurs in field early-type galaxies, but typically with very
modest accretion rates. In contrast, cluster galaxies do not  accrete
cold gas.  All the \HI\ discs that we observed have counterparts of ionised
gas.  Moreover, galaxies with an inner \HI\ disc also have an inner molecular
gas disc.  The strikingly similar kinematics of these different tracers shows
that they are all part of the same structure. The combination of \HI\ and CO
imaging clearly shows, through the detection of gas tails, that the inner
discs are the result of accretion events.  The cold ISM in
the central regions  is dominated by molecular gas
($M_{\rm H_2}/M_{\rm HI} \simeq 10$).

There is no obvious overall relation between current gas content and internal
dynamics. This is not very surprising given that in most galaxies with \HI\ it
is due to the recent accretion of small amounts of gas while the dynamical
characteristics are set over much longer timescales. Within our limited number
statistics, the fastest rotating galaxies all posses \HI\ disks. This would
suggest that an accretion/merger involving a large amount of gas is required
to produce the galaxies most dominated by rotation.  However, the reverse is
not true as \HI\ disks can be present also at intermediate $V/\sigma$. The
similarity between the kinematics of the \HI\ and that of the stars seen in
some of the galaxies with large, regular \HI\ disks, suggests that an
accretion/merger involving a large amount of gas was part of the evolution of
those systems.

We observe a close relationship between gas content and the small amounts of
star formation which occurs in some of the \sauron\ galaxies. Galaxies with
widespread star formation are gas rich and are galaxies that have experienced
a recent accretion event. The radio continuum emission detected in these
galaxies is consistent with the star formation observed. However, as already
noticed by \citet[e.g.][]{Morganti06} and \citet{Serra06}, the relation
between \HI\ and the overall properties of the stellar population is very
complex. The few galaxies with a significant young sub-population and/or star
formation, have inner gas discs. For the remaining galaxies there is no trend
between stellar population and \HI\ properties. Very interestingly, we find a
number early-type galaxies that are very gas rich but are not forming stars
and have not done so for a while as they only have an old stellar population.
One example of such a galaxy is NGC 4278 where the fact that the large \HI\
disc shows very regular kinematics implies that this galaxy has been gas rich
for at least a few Gyr. Despite this, there is no evidence for a young stellar
population in this galaxy.  In addition, we find that the stellar populations
of our field galaxies are typically younger than those in Virgo.  This is
expected because field galaxies are likely to have accreted some gas in the
last few Gyr, while in the Virgo cluster this is not the case.  The difference
in stellar population is reflecting of the differences in accretion history of
cold gas between the two groups of galaxies.

In about 50\% of our sample, we detect a central radio continuum source. In
many galaxies, the continuum emission is due to a radio-loud AGN, but
continuum emission from star formation is also detected in some galaxies.
Galaxies with star formation follow the radio-FIR correlation.  The presence
of radio continuum emission correlates with the \HI\ properties, in the sense
that galaxies with \HI\ in the central region are more likely to be detected
in continuum compared to galaxies without \HI.  Galaxies with \HI\ but in
off-centre structures behave in between.  This trend is mainly due to the star
formation observed in galaxies with central gas reservoirs, and is not related
to AGN fuelling.

In this paper, we have presented a number of interesting trends that suggest
that gas and gas accretion plays a role in the evolution of early-type
galaxies, in particular those found in the field. Although our collection of
\HI\ images is the largest and deepest available for early-type galaxies, the
number is still fairly small and the statistical basis for most of the trends
we find is not strong. Similar data on larger samples will be needed to put
the results presented here on a more solid basis. We are in the process of
collecting deep \HI\ imaging data for a much larger sample of early-type
galaxies \citep[the ATLAS$^{\rm 3D}$ sample, see http://purl.org/atlas3d
][]{Cappellari10,Serra09} that will allow us to further investigate the trends
described here.

\section*{Acknowledgments} We thank the anonymous referee for helpful 
suggestions which improved the paper. 
The Westerbork Synthesis Radio Telescope is
operated by the Netherlands Foundation for Research in Astronomy ASTRON, with
support of NWO.  This research has made use of the NASA/IPAC Extragalactic
Database (NED) which is operated by the Jet Propulsion Laboratory, California
Institute of Technology, under contract with the National Aeronautics and
Space Administration. The Digitized Sky Survey was produced at the Space
Telescope Science Institute under US Government grant NAG W-2166.  MC
acknowledge support from a STFC Advanced Fellowship (PP/D005574/1).

\bibliographystyle{mn2e}

\label{lastpage}

\end{document}